# The Interplay Between Forces, Particle Rearrangements, and Macroscopic Stress Fluctuations in Sheared 2D Granular Media

Kwangmin Lee [a], Ryan C. Hurley [a, b] *

[a] *Department of Mechanical Engineering, Johns Hopkins University, Baltimore, Maryland 21218, USA*
[b] *Hopkins Extreme Materials Institute, Johns Hopkins University, Baltimore, Maryland 21218, USA*

**Abstract**

Recent studies have established correlations between non-affine motion and macroscopic stress fluctuations in sheared granular media. However, a comprehensive examination of the relationship between non-affine motion, macroscopic stress fluctuations, and inter-particle forces remains lacking. We investigated this interplay in simulations of 2D granular media during stick-slip events under plane shear. We found that, during most large slip events, particles with the greatest non-affine motion, as quantified by $D^2_{\min}$, initially coalesce into one or two dominant connected clusters. These clusters coincide with the region exhibiting the greatest instantaneous reduction in inter-particle forces, indicating a significant correlation between inter-particle force fluctuations and particle rearrangements. Furthermore, the magnitude of the greatest non-affine motion within these clusters correlates strongly with the magnitude of macroscopic stress fluctuations during slip events. This correlation increased when the non-affine motion of particles in a neighborhood around the point of greatest non-affine motion was included in the analysis, suggesting that plastic events are best understood as regional rather than point-like occurrences. Our results held for various inter-particle friction coefficients. Our findings suggest that elastoplastic models should consider plastic events as regional rather than point-like and highlight the importance of studying the propagation of particle rearrangements.

E-mail address: rhurley6@jhu.edu

# 1. Introduction

    Plastic deformation has been extensively studied in granular materials because of the prevalence of these materials in industrial and geophysical processes, and in natural hazards such as earthquakes and landslides. Plasticity in granular and other disordered solids remains challenging to predict because the structure of these materials lacks long-range order. This lack of long-range order implies that defects which can act as nucleation sites for plasticity cannot be easily identified. In contrast, defects such as dislocations in crystalline materials are straightforward to identify and are closely related to the initiation and progression of plasticity.

    There are broadly two approaches which have been taken to study plasticity in granular and amorphous materials: an approach which focuses on microscopic dynamics such as particle rearrangements and an approach which focuses on the inter-particle force network. Shear transformation zone (STZ) theories align primarily with the first approach. Argon, Spaepen, and others [1-3] inspired the concept of STZs by explaining plastic deformation in metallic glasses as thermally-activated shear transformations initiated around free volume regions. Falk and Langer [4] formally introduced the notion of STZs as local regions where rearrangements occur. To identify STZs, the same authors developed $D^2_{min}$, which measures how local particle dynamics deviate from affine deformation. Langer [5] established an analytical mean-field elastoplastic model (EPM), introducing STZs as zones experiencing plastic deformation. In this model, a quadrupolar elastic displacement field is produced around an STZ, and a corresponding change in the local stress fields occurs. Picard *et al*. [6-7] advanced this model by introducing an EPM which considers the long-range elastic perturbation induced by a localized plastic event. Nicolas *et al*. [8] reviewed various other EPMs for granular and amorphous solids. None of these EPMs consider the effects of rearrangements on local structure, and all EPMs treat long-range elastic perturbations in an idealized manner. Zhang *et al*. [9] showed that softness, a machine-learned measure of local structure, is strongly correlated with the propensity of a local region to rearrange, emphasizing the importance of incorporating local structure into elastoplastic models. Their findings further indicate that plastic rearrangements can influence nearby particles, leading to short-range particle rearrangement propagation, which plays a key role in the evolution of strain localization and plastic flow. Finally, many studies have employed correlation analysis between macroscopic plasticity and structure to identify indicators which may be used to identify structural defects responsible for plastic flow in amorphous solids [10-18]. All prior studies focusing on microscopic dynamics and particle rearrangements provide a strong foundation for understanding macroscopic plasticity, but a stronger link to inter-particle force fluctuations is needed to understand the fully anisotropic and non-local nature of plastic events.

Taking the approach to plasticity based on inter-particle force networks, Cates *et al*. [19] proposed a simple model to explain the jamming to unjamming transition based on the now well-known concept of force chain buckling. Tordesillas [20] examined unjamming transitions and associated instabilities and particle rearrangements in dense granular systems using discrete element method (DEM) simulations, identifying force chain buckling as a primary source of plastic events. Tordesillas and Muthuswamy [21] developed a multiscale thermo-micromechanical continuum model of dense granular materials considering force chain buckling. Tordesillas and Muthuswamy [22] established an analytical model for force chain buckling and performed a parametric study to determine the effect of particle-scale properties on force chain stability. Using force chains and minimal contact cycles, Tordesillas *et al*. [23] studied the evolving stability of jammed states of specific cluster conformations, *i.e.*, particles forming force chains and minimal contact cycles in dense granular materials. Kramer *et al*. [24] and Bretz *et al*. [25] found that by using the average of local measures, including the persistence diagram of force network, the differential force network (which quantifies changes in the magnitude of inter-particle forces over a given time interval), and measures of broken, mobile, and nonmobile contacts, the time to macroscale slip could be predicted. Gupta *et al.* [26] studied force chain behavior occurring during macroscopic stress fluctuations and found that force chains may either buckle or remain stable during macroscopic stress drops depending on the behavior of the surrounding weak force network.

Although the two approaches of analyzing microscopic dynamics and force networks are complementary, studies connecting the concepts of local rearrangements (non-affine motion), macroscopic stress fluctuations, and changes in inter-particle forces remains limited. Examples of such studies include one by Ma *et al*. [27], which examined the initialization and localization of plastic events on various timescales in sheared granular media. They found a moderate correlation between the magnitude of $D^2_{min}$ and local stress fluctuations. Ma *et al*. [28] and Mei *et al*. [29-30] predicted macroscale stress fluctuations based on the $D^2_{min}$ field using machine learning. They found that regions of elevated $D^2_{min}$ are spatially concentrated and tend to generate large clusters inside the granular system during macroscopic stress drop events. However, many fundamental questions remain unresolved, including what type of spatial information is important for the predictions of macroscopic response, whether there are some mesoscopic structures responsible for varying macroscopic responses, and what causes the different spatial patterns of mesoscopic structures [30].

The primary goal of this work is to investigate the connection between microscopic dynamics, inter-particle force fluctuations, and macroscopic stress response in 2D sheared granular media. To achieve this, we employed 2D DEM simulations of granular materials subjected to plane shear. We characterized representative stick and slip events based on the macroscopic response and analyzed microscopic dynamics and the differential force network to better understand the microscopic rearrangements and force fluctuations around slip events. Additionally, we performed statistical analyses to determine whether the observed characteristics of these representative stick and slip events are frequently observed in the stick-slip regime. Finally, we conducted correlation analyses to explore the relationship between microscopic dynamics and macroscopic stress fluctuations.

The remainder of the paper is structured as follows. Section 2 introduces the DEM model used for plane shear simulations. Section 3.1 discusses the macroscale DEM results for a sample containing around 5,000 particles. The relationship between the microscopic dynamics and the differential force network for this sample is analyzed in detail in Section 3.2. Section 3.3 presents a statistical analysis of stick-slip events and non-affine motion, and a correlation analysis between the macroscale response and various kinematic measures. Section 3.4 discusses the expansion of DEM simulations to other inter-particle friction coefficients. Section 4 offers a discussion, and Section 5 provides concluding remarks.

## 2. Discrete Element Method (DEM) Model

We created a two-dimensional DEM model of a plane shear test (Fig. 1) in the open-source code LIGGGHTS, version 3.8.0 [31]. We used a uniform distribution of particle radii between 20-40 $\mu$m. The number of particles in the simulation was approximately 5,000. Top and bottom walls were formed using particles with a fixed radius of 40 $\mu$m and were initialized on horizontal lines (i.e., in the $x$ direction), as shown in Fig. 1. The particles constituting the bottom wall were fixed, while the particles constituting the top wall were constrained to move together. Boundaries in the +/-$x$ direction were periodic. The simulation consisted of three steps: (i) the initial positions of particles were generated using MATLAB's rand function and placed between the top and bottom walls. The first particle was placed randomly within the gap between the walls. For each subsequent particle, a random position was generated, and the particle was accepted only if its center was at least 80 $\mu$m away from all previously placed particles. Otherwise, the position was discarded and re-sampled. This process was repeated until 5,000 particles were successfully placed; (ii) the top wall was moved down (in the -$y$ direction, as shown in Fig. 1) until a constant normal force, $P$, was reached; (iii) the top wall was moved horizontally, with horizontal velocity, $V$. In both steps, the $y$ positions of particles constituting the top wall were integrated using a velocity-Verlet scheme to maintain a constant normal force, $P$. The mass of the top wall used in the velocity-Verlet scheme was chosen to be the sum of the masses of the particles that constituted it. Particles between the top and bottom walls constituted the sample and were integrated using a velocity-Verlet scheme, and gravity was neglected in all calculations. The sample's normal force, $P$, and shear force, $S$, are the sum of the normal and shear forces, respectively, acting on the particles constituting the top wall.

To ensure that the sample was flowing in the quasi-static regime, the horizontal velocity $V$ was set to maintain an inertial number $I$ of $1 \times 10^{-6}$ [32]. The inertial number for a 2D simulation is defined as [32]

$$I = \dot{\gamma}\sqrt{\frac{m}{p}}, \tag{1}$$

where $\dot{\gamma}$ is the macroscale shear rate, defined as $V/H$ and equal to 0.1 s$^{-1}$, $m$ is the mean mass of the particles, and $p$ is the normal pressure, defined as $P/W$. The parameters $H$ and $W$ represent the height and width of the sample, respectively, as shown in Fig. 1. $H$ was measured immediately after step (ii) of the simulation, during which the normal force, $P$, reached a constant value.

For the normal and tangential contact models, the Hertzian contact model for cylindrical bodies [33] and Lai *et al.*'s semi-analytical Hertzian frictional model in two dimensions [34] were used, respectively, both incorporating the linear damping models [35]. For the rolling resistance model, the type C elastic-plastic spring-dashpot model [36] was employed. Appendix A provides a detailed description of the contact models.

To ensure that particle contacts remained in the elastic regime, we measured the mean normalized overlap $<\delta_n/d>$, where $\delta_n$ is the normal overlap displacement and $d$ is the particle diameter. After step (ii), we found that $<\delta_n/d> \approx 0.25\%$. According to Ref. [37], values below 1% are sufficient to ensure that variations in stiffness have negligible influence on DEM simulation outcomes. These observations confirmed that our system operated within a regime where particle stiffness effects are negligible. We also confirmed that almost all contacts after step (ii) satisfied $\delta_n/d < 1\%$, with only a small fraction (0.2%) falling in the slightly higher range of $1\% < \delta_n/d < 1.5\%$.

For material properties, 1 GPa, 0.2, 0.9, and 2500 kg/m$^3$ were chosen for Young's modulus, Poisson's ratio, coefficient of restitution, and density, respectively. For the sliding friction coefficient, $\mu_s$, and the rolling friction coefficient, $\mu_r$, 0.7 and 0.01 were used, respectively. Here, we assumed that the particle shape was circular, and the actual contact moment was negligible, and thus a small value was chosen for $\mu_r$. The contact moment introduced in this DEM model is an artificial means of capturing the effect of particle shape irregularities. Simulations for various values of $\mu_s$ were also conducted, and the results are provided in Section 3.4. Varying $\mu_s$ did not qualitatively change our results. The normal and tangential force damping ratio, $\alpha$, and the rolling viscous damping ratio, $\eta_r$, were set to 0.1 and 0.3, respectively. The viscoelastic damping constant for the normal contact force, $\gamma_n$, calculated from Eq. (A4), is approximately 5 N·s/m. Since the simulation is quasi-static, we confirmed that the normal damping force, $F_n^d$, was negligible; its magnitude was approximately 0.1% of the corresponding normal elastic contact force, $F_n^e$, after step (ii).

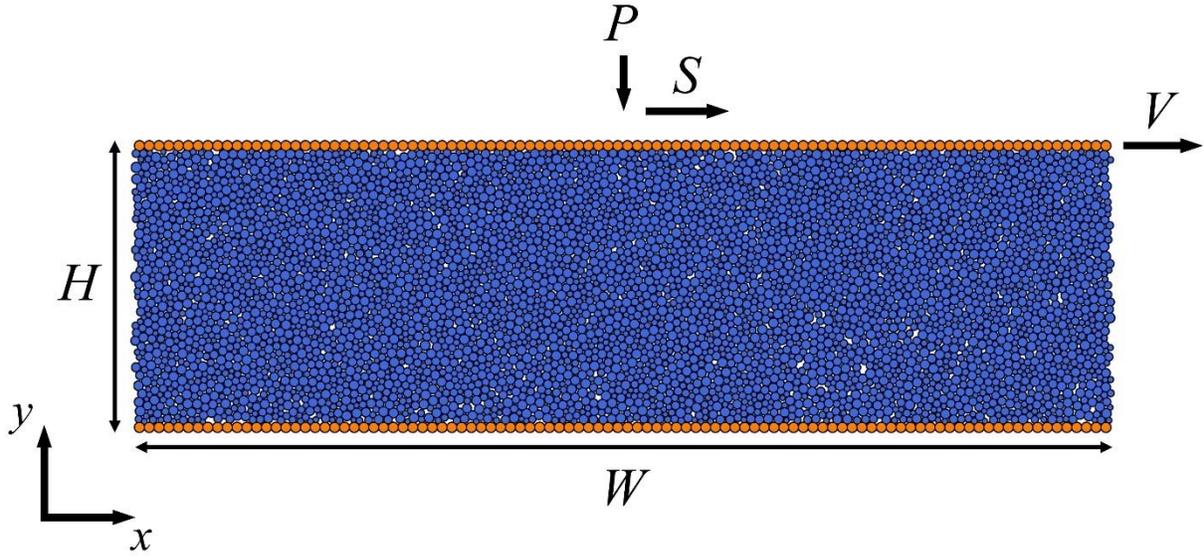

Fig. 1. DEM model for a plane shear test.

## 3. DEM Results

3.1. Macroscopic response

The nondimensional shear stress versus strain curve is shown in Fig. 2. The nondimensional shear stress, $\tau$, is defined as the ratio of shear to normal force, $S / P$, where both $S$ and $P$ are measured in the sheared granular material, as described in Section 2.

We focused our analysis on stress fluctuation events occurring after 3% macroscopic shear strain. We refer to this portion of the simulation as the stick-slip regime because the sample continued to deform at approximately constant stress with intermittent stick-slip events, as shown in Fig. 2. In the stick-slip regime, we observed approximately 25,000 stick and slip events. The volume fraction of the sample continued to increase during this regime, rising by approximately 1% throughout the stick-slip regime. This indicated that the system had not yet reached the critical state. A stick event begins when $\tau$ increases and ends when this increase stops. A slip event begins when $\tau$ decreases and ends when this decrease stops.

To gain insight into the nature of slip events, we first focused on the region near the largest slip event in the stick-slip state regime, specifically between 12.882% and 12.885% shear strain. Figure 3a shows the nondimensional shear stress-strain curve in this region. At the beginning of the slip event, $\tau$ began to decrease gradually but remained almost constant up to step 2 (as marked in Fig. 3a). Then, a significant drop occurred. After the slip event, $\tau$ exhibited fluctuations. While these fluctuations are clearly visible in Fig. 3a, their gradual convergence is not fully captured within this strain range, but occurs at later strains due to energy dissipation and viscous damping. Figure 3b presents the normal and shear forces over the same strain range. The normal force, $P$, remained nearly constant, while the shear force, $S$, exhibited a sharp drop during the slip event. Figure 3c illustrates the evolution of the sample height, $H$, which decreased during the slip event, indicating compaction. After the slip event, both $\tau$ and $H$ fluctuated due to stress redistribution and ongoing particle rearrangement but gradually stabilized over time.

In addition to slip events, we also studied the nature of stick events. The largest stick event immediately followed one of the slip events and thus did not exhibit the typical characteristics of a stick event. Since this event was directly influenced by the preceding slip event, we instead analyzed the second-largest stick event, which exhibited the characteristic features of stick events commonly observed in our analysis (see Fig. 4).

Figure 4a shows the nondimensional shear stress-strain curve around the stick event between 13.414% and 13.518% shear strain. In this stick event, $\tau$ increased gradually over time. Figure 4b presents the normal and shear forces over the same strain range. Here, the normal force ($P$) remained nearly constant, while the shear force ($S$) increased gradually during the stick event. Figure 4c illustrates the evolution of the sample height ($H$), which increased during the stick event, indicating dilation.

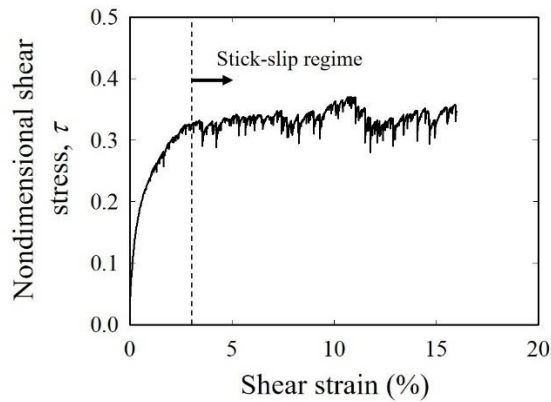

Fig. 2. Nondimensional shear stress-strain curve from discrete element method simulation.

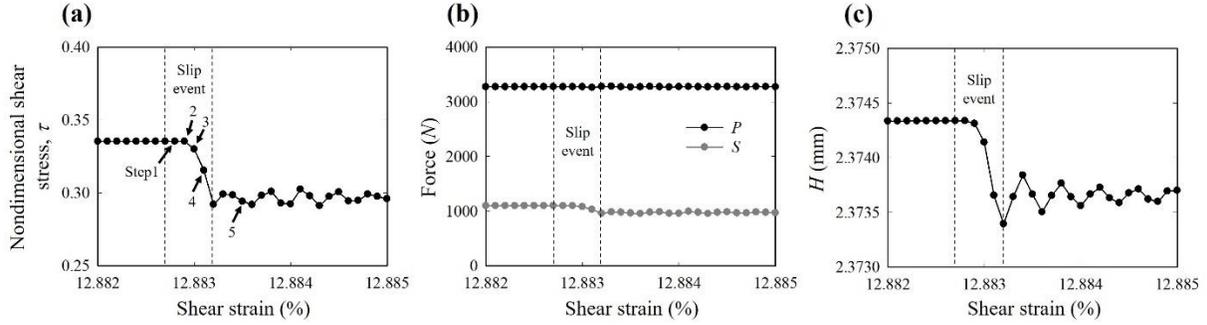

Fig. 3. (*a*) Nondimensional shear stress-strain curve around a slip event occurring between 12.882% and 12.885% shear strain, (*b*) Normal force (*P*) and shear force (*S*) within the same strain range, and (c) Evolution of the sample height (*H*) within the same strain range.

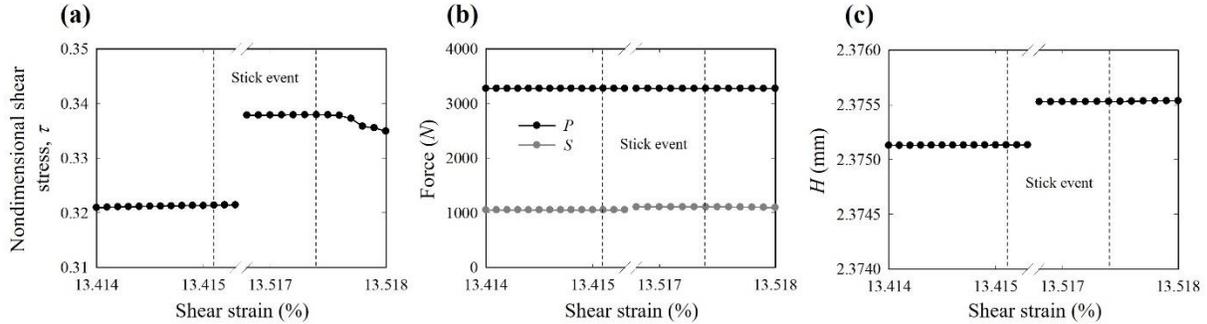

Fig. 4. (*a*) Nondimensional shear stress-strain curve around a stick event occurring between 13.414% and 13.518% shear strain, (*b*) Normal force (*P*) and shear force (*S*) within the same strain range, and (c) Evolution of the sample height (*H*) within the same strain range. A portion of the shear strain axis is omitted (indicated by the double slashes in all panels) to highlight the stick event interval. The apparent step-like changes in (a)-(c) are visual artifacts caused by this omission and do not represent discontinuities in the data.

3.2. Microscopic dynamics and differential force network analysis around a slip event

To understand the kinematic and kinetic behavior at the particle scale around the largest slip event shown in Figure 3, we examined the differential force network, the $D^2_{min}$ field, the non-affine displacement field, and the shear strain field. The macroscale shear strain increments between which inter-particle forces and particle positions were output from the DEM simulation, $\Delta\gamma$, was $1 \times 10^{-6}$ for this analysis. A statistical analysis of the kinematic and kinetic behavior around all slip events is provided in Section 3.3.

Figure 5 shows the 1% of particles exhibiting the greatest reduction in inter-particle force magnitude and the differential force network at each $\Delta\gamma$ increment at the five timesteps indicated in Figure 3a. The differential force network was calculated as $f^\alpha_t - f^\alpha_{t-1}$ of each contact $\alpha$, where $f^\alpha_t$ and $f^\alpha_{t-1}$ are the magnitudes of inter-particle force vectors at the current time step and previous time step (separated by a strain increment of $1 \times 10^{-6}$), respectively. If force-bearing contacts at the previous time step are created or destroyed at a given time step, the differential force network at this contact is set to zero in the current time step. As shown in Figure 5a, differential forces were randomly distributed at the onset of the slip event. When the significant stress drop began, the locations of the most-negative differential forces concentrated towards a single region in the middle-left portion of the sample and formed a single cluster (Fig. 5b). To further examine this phenomenon, we introduce the concept of a GR region, which refers to the 1% of particles exhibiting the greatest reduction in inter-particle force magnitudes. A single clustered GR region was first established at the onset of the significant stress drop (Fig. 5b), as reflected through quantitative analysis using the DBSCAN (Density-Based Spatial Clustering of Applications with Noise) clustering method in MATLAB R2020a [38]. As the stress drop progressed, the single clustered GR region no longer existed (Fig. 5d). Fig. 6 shows close-up views of the area highlighted by a black-bordered rectangle in the differential force network in Fig. 5b-d (step 2, step 3, and step 4 in Fig. 3a). Figure 6 shows that differential force changes did not remain localized. Instead, the changes propagated outward from the initially clustered GR region. In any contact points other than those in the GR region, inter-particle force change was almost zero during steps 2-4 (Fig. 5). Thus, a macroscale stress drop appears to occur due to a significant inter-particle force decrease in the GR region and the subsequent structural change in the sample. After the slip event ended, the initially clustered GR region dispersed and the differential forces became randomly distributed again (Fig. 5e).

Figure 7 illustrates the 1% of particles exhibiting the greatest $D^2_{min}$, and the full $D^2_{min}$ field at each $\Delta\gamma$ increment at the five timesteps indicated in Figure 3a. $D^2_{min}$ was computed based on the definition in [30] with a local averaging region of cutoff radius $r = 5r_p$, where $r_p$ is the mean radius of particles. The details of the $D^2_{min}$ computation can be found in Appendix B. In theory, an STZ can be considered a local plastic event; on the other hand, a group of rattler particles undergoing a change in configuration may be identified as an STZ using $D^2_{min}$. We define rattler particles as particles inside the sample that have two or fewer inter-particle contacts, and particles in the top or bottom walls that have one or fewer inter-particle contacts. Because these rattler particles are not involved in stress transmission during a plastic event, they were excluded from the computation of $D^2_{min}$. The colorless disks in the $D^2_{min}$ graphs in the right column of Figure 7 represent the rattler particles.

As shown in Figure 7a, $D^2_{min}$ values were also somewhat randomly distributed spatially at the onset of the slip event, similar to the differential force network (Fig. 5a). When the significant stress drop began, locations with high $D^2_{min}$ values also concentrated towards a single region in the left part of the sample and formed a single cluster (Fig. 7b). To further examine this phenomenon, we introduce the concept of a $GD^2_{min}$ region, which refers to a set of particles that have the greatest 1% of all $D^2_{min}$ values. At the onset of the significant stress drop, a single clustered $GD^2_{min}$ region emerged (Fig. 7b), as identified through DBSCAN clustering analysis (see Fig. 10c for details). This single clustered $GD^2_{min}$ region coincided spatially and temporally with the single clustered GR region (Figs. 5b and 7b), suggesting a strong correlation between interparticle force reduction and particle rearrangement (see Fig. 10a for details). During the significant stress drop, particle rearrangement and nonaffine motion propagated outward from the initially clustered $GD^2_{min}$ region, similar to the evolution of the differential force network (Fig. 7b-d). After the slip event ended, the initially clustered $GD^2_{min}$ region dispersed and the $D^2_{min}$ values became randomly distributed (Fig. 7e).

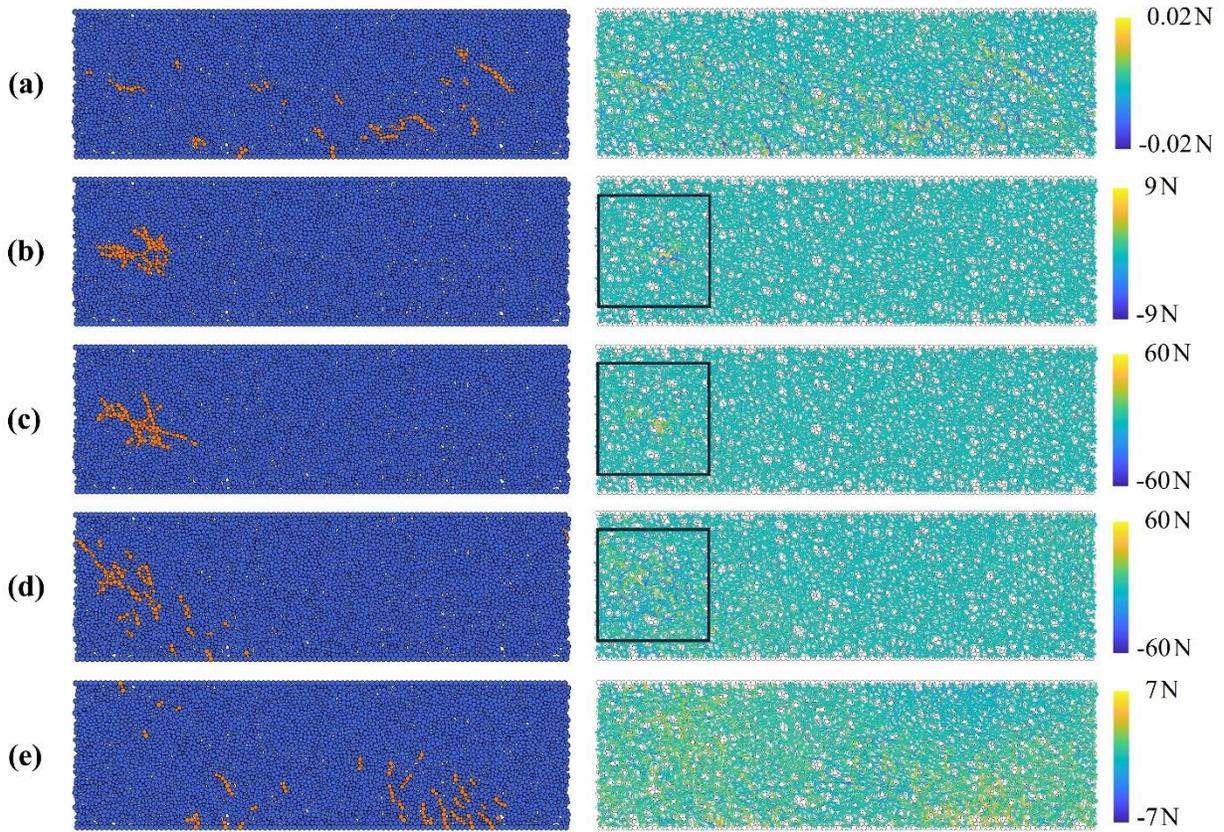

Fig. 5. The particles exhibiting the greatest 1% reduction in inter-particle force magnitude (left column) and the differential force network (right column) at (a) step 1, (b) step 2, (c) step 3, (d) step 4, and (e) step 5, as indicated in Fig. 3a. The boxed regions in the right column are shown in more detail in Fig. 6.

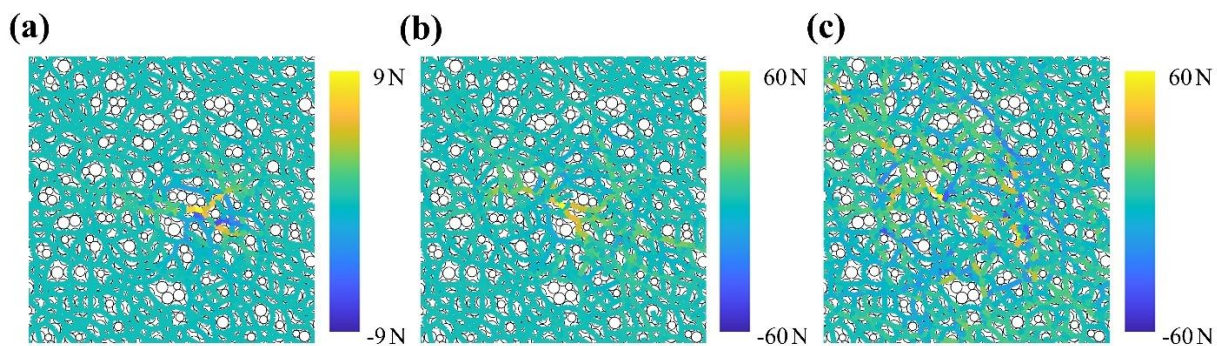

Fig. 6. Close-up views of the area highlighted by a black-bordered rectangle in Fig. 5b-d, shown at (a) step 2, (b) step 3, and (c) step 4, as indicated in Fig. 3a.

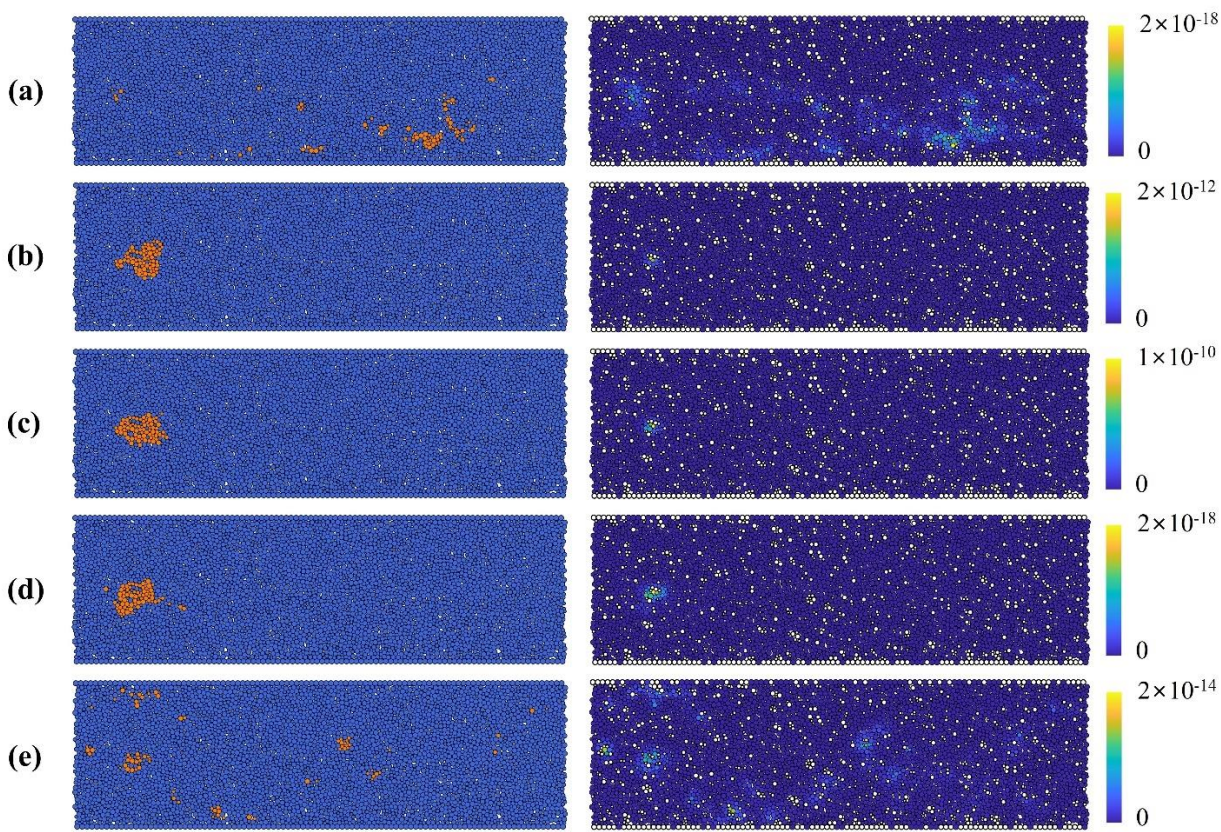

Fig. 7. The 1% of particles exhibiting the greatest $D^2_{min}$ (left column) and the full $D^2_{min}$ field (right column) at (a) step 1, (b) step 2, (c) step 3, (d) step 4, and (e) step 5, as indicated in Fig. 3a.

Figure 8 illustrates the non-affine displacement field during the significant stress drop, obtained by subtracting the global affine displacement field from the total displacement field. At each of steps 2–4 (Fig.

8a–c), a localized region of high non-affine displacement was observed, which spatially overlapped with the single clustered region of high $D^2_{min}$ shown in Fig. 7b-d. During these steps, the non-affine displacement field exhibited spatially correlated structures. In step 2 (Fig. 8a), the displacement vectors loosely resembled a quadrupolar pattern, which has been noted in earlier studies of amorphous solids [39-40]. However, in steps 3 and 4 (Fig. 8b-c), the patterns were more irregular and did not clearly conform to any specific symmetry. These observations suggest that while quadrupolar-like structures may appear transiently, they are not a universal feature of the non-affine response during stress drops.

To further investigate the dominant symmetry of the deformation field around the STZ during the significant stress drop, we analyzed the shear strain field instead of the non-affine displacement field. According to Ref. [41], the STZ exhibits different shear strain field symmetries depending on its dominant deformation mode. It primarily exhibits quadrupolar symmetry under pure shear and tension/compression, dipolar symmetry under torsion, and both quadrupolar and dipolar symmetry under simple shear. These shear strain field symmetries result from elastic propagation, as described by Eshelby inclusion theory. This theory explains how a local shear transformation generates a long-range strain field in an elastic medium.

We define the shear strain to be the *xy* component of the granular Lagrangian strain, whose computation and shear strain field construction are detailed in Appendix C. The left column in Figure 9 shows the shear strain field during the significant stress drop. By comparing Figures 7 and 9, we confirm that the STZ, where $D^2_{min}$ is the greatest, coincides with the location where the shear strain is the greatest. A Fast Fourier Transform (FFT) was performed to analyze the angular wavenumber spectrum along dashed circles surrounding the location where the shear strain is greatest in the sample. Theoretically, an STZ has been considered a point-like event, suggesting that the choice of a larger or smaller radius for the dashed circles should not significantly affect the Fourier analysis.

However, careful selection of the radius is necessary, as excessively large or small values can influence the results. As discussed in Ref. [41], increasing the radius rapidly weakens the characteristic patterns of the STZ because spatial resolution decreases in proportion to the inverse square of the radius. Conversely, if the radius is too small, the selected region may lie near the center of the STZ, which is dominated by positive shear strain. In either case, the analysis can deviate from the idealization of STZs as point-like events and obscure relevant characteristics [41]. Based on these considerations, we chose the radius of the dashed circles as $8r_p$ in this study. The Fourier spectrum for different dashed circle radii at the same steps is presented in the Supplemental Materials (Section S1) [42] to illustrate the effect of radius on the Fourier spectrum results. The right column in Figure 9 presents the Fourier spectrum of angular wavenumbers, computed at discrete values from 2 to 8, along the dashed circle surrounding the location where the shear strain is greatest. A wavenumber of 4 indicates a quadrupolar response. A wavenumber of 2 indicates a dipolar response. Higher wavenumbers suggest more complex non-dipolar and non-quadrupolar responses.

As shown in Fig. 9a, at the onset of the significant stress drop, the FFT spectrum shows that wavenumber 4 has a slightly higher magnitude than others. However, the overall differences are modest, and several other wavenumbers have comparable values, indicating that the shear strain field contains only a weak or ambiguous quadrupolar component rather than a distinct quadrupolar pattern. As the stress drop progresses (Figs. 9b–c), the spectrum becomes even more evenly distributed across wavenumbers, suggesting that no particular angular mode dominates. This supports the view that the strain field exhibits a broad and irregular spectrum rather than a consistent dipolar or quadrupolar symmetry.

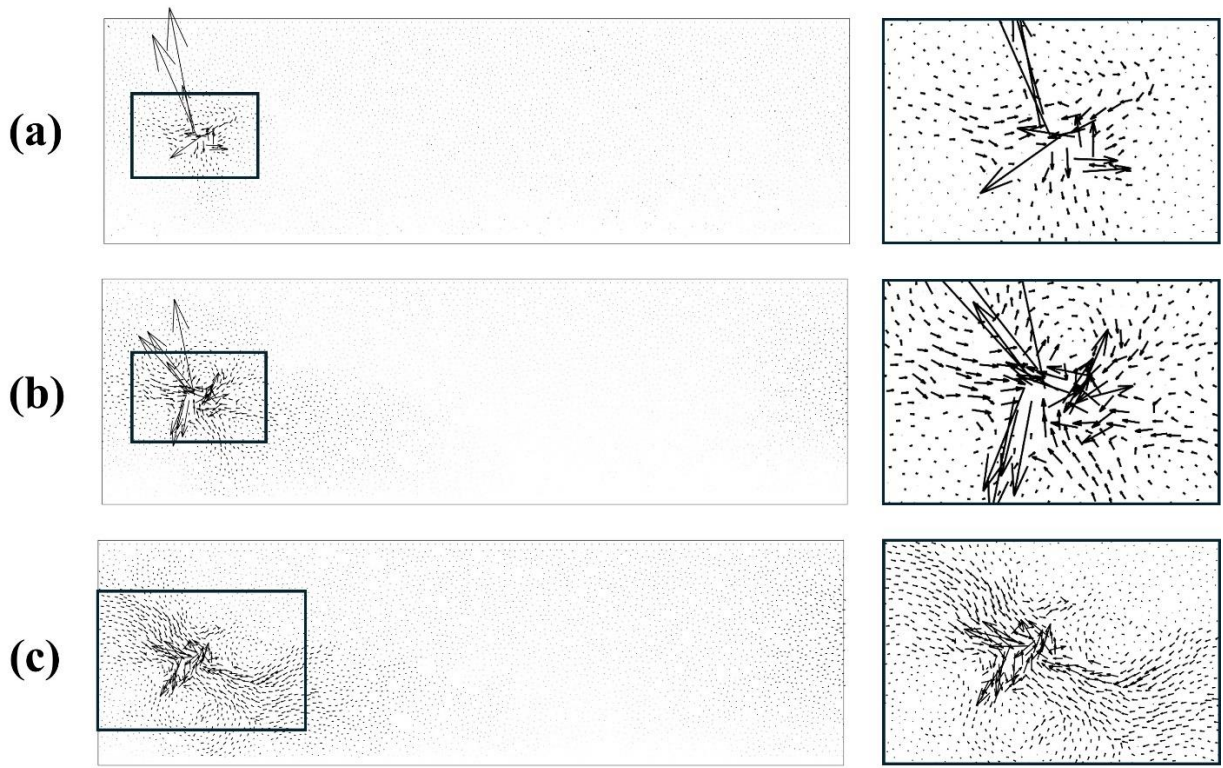

Fig. 8. Non-affine displacement field (left column) and close-up views of the area highlighted by a black-bordered rectangle in the non-affine displacement field (right column) during the significant stress drop at (a) step 2, (b) step 3, and (c) step 4, as indicated in Fig. 3a. The arrows represent displacement vectors, where the arrow length indicates relative displacement magnitude. Note that the arrow lengths are scaled differently for each step to highlight relative displacements.

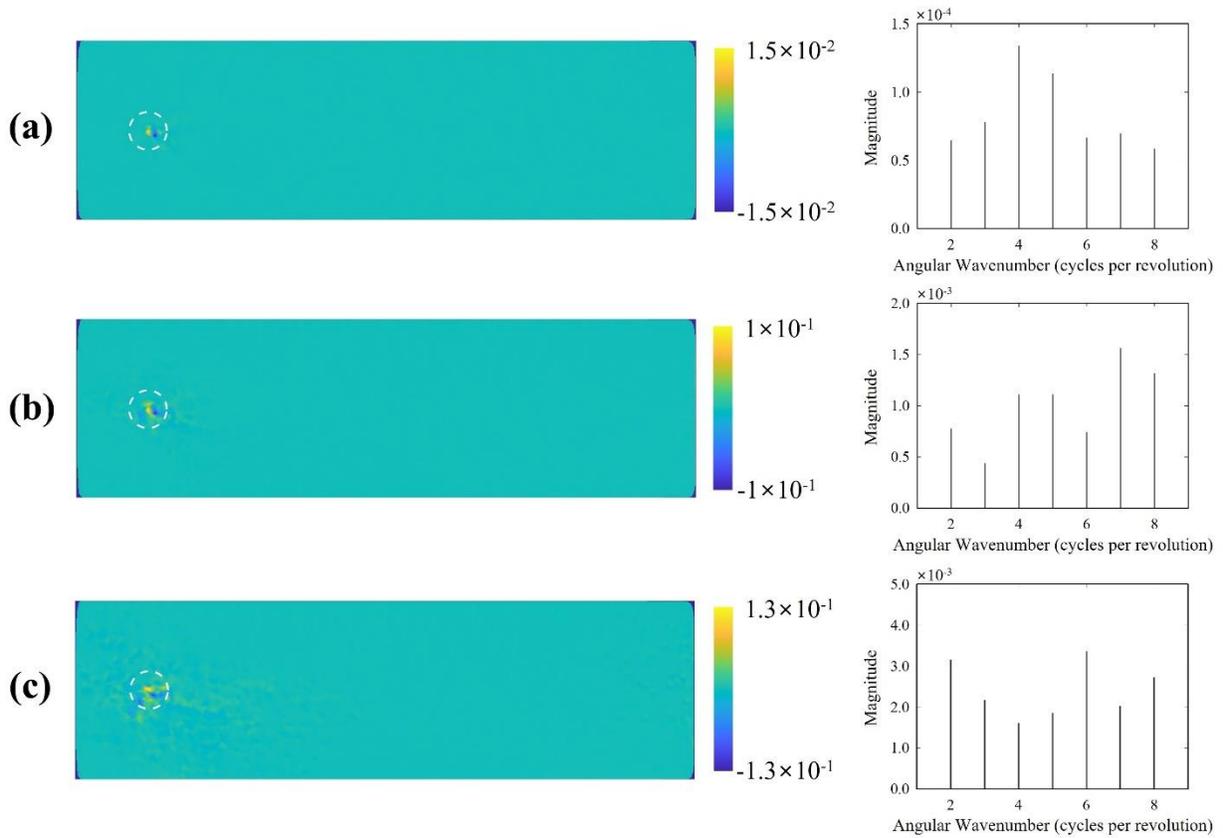

Fig. 9. Shear strain field (left column) and Fourier spectrum of angular wavenumbers (right column), computed at discrete values from 2 to 8 along dashed circles centered at shear transformation zones at different time steps during the significant stress drop: (a) step 2, (b) step 3, and (c) step 4, as indicated in Fig. 3a.

Figure 10 presents a detailed analysis of kinematics and kinetics during the slip event between 12.882 and 12.885% sample shear strain. It includes calculations of the correlation coefficient between stress changes and non-affine motion at particles (Fig. 10a), the greatest $D^2_{min}$ value and the mean of the 10% greatest $D^2_{min}$ values (Fig. 10b), the number of clusters (calculated as described in subsequent paragraphs) in the GR region and the $GD^2_{min}$ region, respectively (Fig. 10c), and the mean and median absolute values of contact angles in the GR region (Fig. 10d).

The Spearman correlation coefficients between the absolute values of the maximum principal stress changes, $|\Delta\sigma_p|$, and $D^2_{min}$ of each particle in this strain range are shown in Figure 10a. This curve shows that the correlation between $|\Delta\sigma_p|$ and $D^2_{min}$ was highest at step 3 when the single clustered $GD^2_{min}$ region spatially and temporally coincided with the single clustered GR region (Figs. 5c and 7c). However, from

step 4, the correlation between $|\Delta\sigma_p|$ and $D^2_{min}$ became weak, following the collapse of the single clustered GR region and the propagation of particle rearrangements.

Figure 10b illustrates the greatest $D^2_{min}$ value, denoted as $gD^2_{min}$, and the mean of the greatest 10% $D^2_{min}$ values, denoted as $g_{t10\%}D^2_{min}$, within the sample across each timestep. Defining a stable structure as one in which particle rearrangement does not occur, the sample exhibited stability at the beginning of the slip event, marked by low $gD^2_{min}$ values. At the onset of the significant stress drop, which occurs at step 2, we observed an increase in $gD^2_{min}$, which signifies the beginning of an instability. After the slip event ended, $gD^2_{min}$ gradually decreased, indicating a return toward stability. The $g_{t10\%}D^2_{min}$ shows the same trend with $gD^2_{min}$ during this strain range, with slightly lower values.

Figure 10c illustrates the number of clusters in the GR region and the $GD^2_{min}$ region, respectively, within this strain range. To identify and differentiate these clusters, the DBSCAN algorithm was employed in MATLAB R2020a [38]. The algorithm was applied to the positions of the particles with the greatest 1% reduction in inter-particle force for the GR region and to the positions of particles exhibiting the greatest 1% $D^2_{min}$ for the $GD^2_{min}$ region. The parameter set for each cluster included a neighborhood search radius of $r = 5r_p$ and a minimum requirement of one neighbor for a point to be considered a core point. Initially, at the beginning of the slip event, the clusters in both the GR and $GD^2_{min}$ regions were numerous and spatially dispersed, as visually observed earlier in Figures 5 and 7. At the onset of the significant stress drop, a single cluster emerged from the DBSCAN analysis, which later dispersed. The number of clusters in both regions fluctuated after the slip event.

Figure 10d shows the mean and median of the absolute values of the contact angles in the GR region over the same strain range. The contact angle at a contact point $\alpha$, $\theta_\alpha$, is defined as an angle (in radians) between the $x$-axis and an inter-particle force vector. At the onset of the slip event, both the mean and the median of the absolute values of contact angles in the GR region remained unchanged, possibly because no significant particle rearrangement occurred, allowing the structure to remain stable. On the other hand, during the significant stress drop, they were no longer constant, as significant particle rearrangement took place, and the single clustered GR region was not sustained and eventually collapsed. After the slip event, they fluctuated, and these fluctuations gradually decreased and eventually stabilized during a long stick event, even though this is not presented within the strain range shown in Figure 10d. Therefore, these fluctuations likely occurred because the sample remained unstable for some time after the slip event, with particle rearrangement continuing at randomly distributed locations.

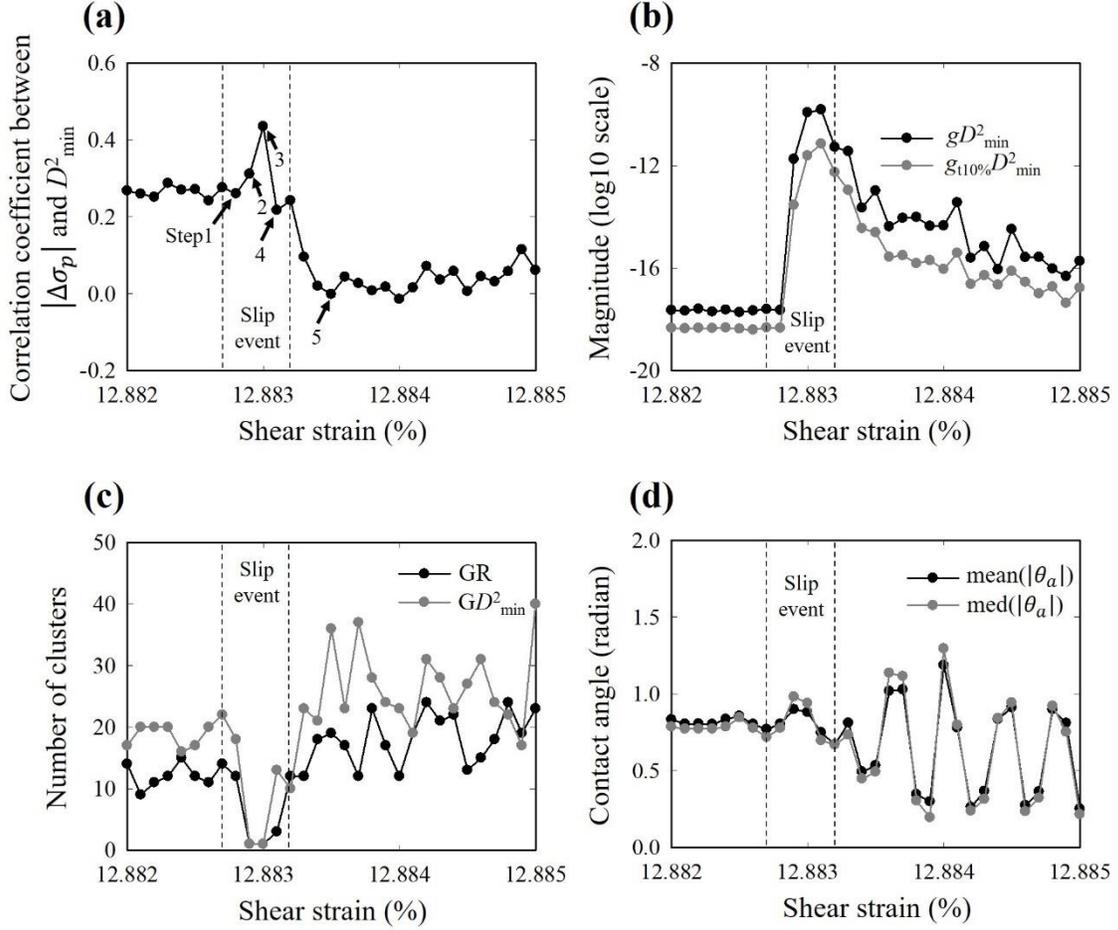

Fig. 10. (*a*) The Spearman correlation coefficients between the absolute values of maximum principal stress change, $|\Delta\sigma_p|$, and $D^2_{min}$ at particles, (*b*) the greatest $D^2_{min}$ and the mean of the greatest 10% $D^2_{min}$ magnitudes, (*c*) the number of clusters in the GR region and the $GD^2_{min}$ region, respectively, and (d) the mean and the median of the absolute values of the contact angles in the GR region during 12.882 to 12.885% sample shear strain.

3.3. Statistical and correlation analysis of stick-slip events

In Sections 3.1–3.2, we identified the representative characteristics of stick and slip events in the stick-slip regime through detailed analysis of the selected events. In addition, we analyzed the microscopic dynamics and the differential force network in detail around the representative slip event. In this section, using data from approximately 130,000 timesteps and 25,000 observed stick and slip events for shear strain ≥ 3% in the stick-slip regime, we conducted a statistical analysis to determine whether these representative characteristics are consistently observed across multiple events, thereby confirming them as typical characteristics of stick and slip events. We also examined the statistical behavior of non-affine motion in greater detail (Section 3.3.1). Then, we investigated correlations between macroscale response and various kinematic measures (Section 3.3.2).

3.3.1. Statistical analysis of stick-slip events and non-affine motion

In Section 3.2, we discovered that during the studied stress drop, particles with the greatest reduction in inter-particle forces (GR region) and those exhibiting high non-affine motion ($GD^2_{min}$ region) were spatially clustered. This single-clustered $GD^2_{min}$ region coincided with the single-clustered GR region (Figs. 5b and 7b), from which particle rearrangements propagated. During the stress drop, a weak quadrupolar characteristic of the shear strain field was observed. However, as particle propagation occurred, the shear strain field did not strictly conform to a discrete quadrupolar or dipolar pattern.

Building on these observations, we now investigate whether the clustering behavior in the GR and $GD^2_{min}$ regions is consistently observed across all slip events. We also examine whether the time duration of a stick event tends to increase with the macroscale cumulative shear stress change during the event. Additionally, we analyze whether the STZ tends to exhibit quadrupolar or dipolar characteristics across slip events.

Figure 11a shows the number of stick and slip events within each range of $\Delta \tau_p$, the cumulative nondimensional shear stress change from the start to the end of a stick or slip event, considering only events with a time duration ($t_p$) greater than $3\Delta\gamma$. We define such events as "long events." Events with $t_p \leq 3\Delta\gamma$ are defined as "short events" and are excluded from the analysis, as most of them (approximately 90%) correspond to transient stress fluctuations following slip events, although the numerical results are not shown. Figures 11b and 11c show the percentage of the minimum number of clusters during the long events in the stick-slip regime for different bins of $\Delta \tau_p$, in the GR and $GD^2_{min}$ regions, respectively.

To analyze whether this clustering is consistently observed during long slip events, we examined how frequently the minimum number of clusters observed in both GR and $GD^2_{min}$ regions is 1 or 2. Figure 11b-c indicate that, during the large slip events ($-10^{-1} < \Delta \tau_p < -10^{-3}$), the percentage of events where the minimum number of clusters in both regions is 1 or 2 is very high, confirming the observed clustering of particles with the greatest reduction in differential force and those exhibiting high $D^2_{min}$. Furthermore, when the magnitude of $\Delta \tau_p$ decreases ($-10^{-3} < \Delta \tau_p < -10^{-4}$), the percentage of slip events where the minimum number of clusters is 1 or 2 in both regions decreases. However, despite this decrease, the clustering percentage remains relatively high.

In contrast, clustering is rarely observed during the long stick events ($\Delta \tau_p > 0$). However, an exception appears for the large stick events ($10^{-2} < \Delta \tau_p < 10^{-1}$). Figure 11a shows that this range contains only 8 stick events; therefore, the apparent increase in the percentage of events where the minimum number of clusters in both regions is 1 or 2 should be interpreted with caution. Analyzing these 8 stick events, we found that most of these 1 or 2 clusters appeared at the end of the stick event, immediately before a slip event. Thus, this clustering may be associated with the slip event rather than being a typical characteristic of large stick events.

Figures 11d and 11e show two-dimensional histograms of $\Delta\tau_p$ and $t_p$ for the long stick events and the long slip events, respectively, during the stick-slip regime. The color represents the number of events in each bin. Figure 11d indicates a positive correlation between $\Delta\tau_p$ and $t_p$ during the long stick events. In contrast, Figure 11e shows no clear correlation between $\Delta\tau_p$ and $t_p$ for the long slip events. This correlation may stem from the fact that, during long stick events, the structure of the sample remains stable and features small $gD^2_{min}$ values. As a result, $\tau$ increases gradually in a more sustained manner. This structural stability likely contributes to the positive correlation observed between $\Delta\tau_p$ and $t_p$ during long stick events.

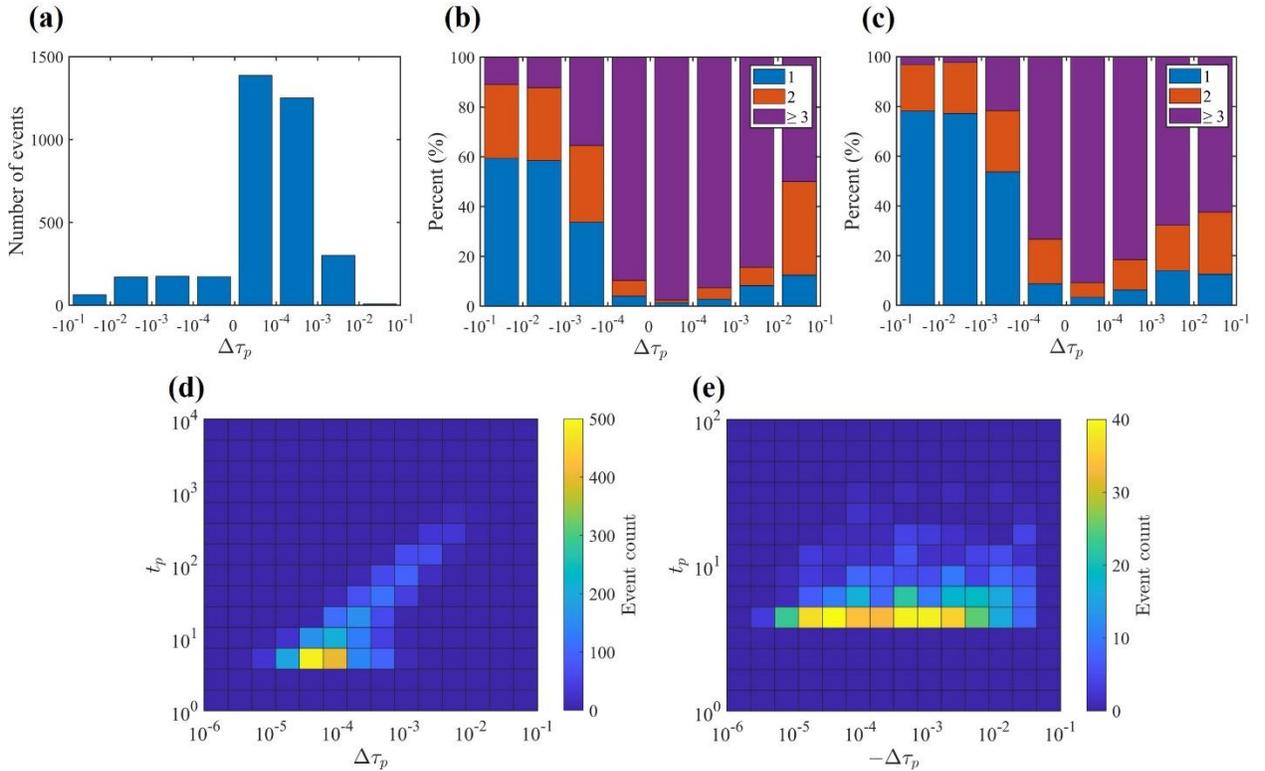

Fig. 11. Statistics of the long events during the stick-slip regime. (a) The number of stick and slip events within each range of $\Delta\tau_p$. (b, c) The percentage of the minimum number of clusters during an event for different bins of $\Delta\tau_p$, in the GR region for (b) and the $GD^2_{min}$ region for (c). (d, e) Two-dimensional histograms showing the distribution of $\Delta\tau_p$ and $t_p$ for (d) the stick events and (e) the slip events. The color represents the number of events in each bin.

To examine the statistical behavior of non-affine motion in detail, we performed FFT along a dashed circle with a radius of $8r_p$ surrounding the location where the shear strain is greatest, as shown in Figure 9, for the long events during the stick-slip regime. In this statistical analysis, approximately 40% of the long events in the stick-slip regime were excluded because the locations where the shear strain is greatest were near the sample boundary, causing the dashed circle to extend beyond the sample boundary.

Figure 12 illustrates the percentage of events in which the Fourier spectrum (computed at discrete wavenumbers from 2 to 8) has the highest value at wavenumbers 4, 2, or others within each range of $\Delta\tau_p$ for the long events during the stick-slip regime. The results are shown for (a) the first, (b) the second, and (c) the third timesteps, described next. The first timestep corresponds to the moment when the minimum number of clusters is first observed in the $GD^2_{min}$ region during an event, as exemplified by step 2 in Figure 10c. The third timestep represents the point at which $g_{t10\%}D^2_{min}$ reaches its maximum during an event, as exemplified by step 4 in Figure 10b. The second timestep is the midpoint between the first and third timesteps within an event, as exemplified by step 3 in Figure 10a.

Figure 12 shows that the percentage of events in which the Fourier spectrum has the highest magnitude at wavenumber 4 or 2 remains low across the first, second, and third timesteps within each range of $\Delta\tau_p$. These percentages are close to what would be expected if a wavenumber were randomly selected from the seven discrete bins between 2 and 8, indicating no statistically dominant quadrupolar or dipolar symmetry.

The lower percentage of events exhibiting quadrupolar or dipolar symmetry may also be influenced by the calculation method for $D^2_{min}$. Previous studies [28, 30] have computed $D^2_{min}$ over a longer time interval, such as an entire slip or stick event, from its initiation to its termination. In contrast, our analysis examines $D^2_{min}$ over smaller and fixed time intervals within the stick and slip events, which provides finer temporal resolution but may also capture localized fluctuations that obscure the overall quadrupolar or dipolar symmetry. When $D^2_{min}$ is integrated over an entire event, these local variations may average out, potentially enhancing the visibility of quadrupolar or dipolar symmetry. To assess this possibility, we recalculated $D^2_{min}$ using the conventional approach—integrating over an entire event—to determine whether a more prominent quadrupolar or dipolar symmetry emerges. The results of this comparison are provided in the Supplemental Materials (Section S2) [42] and do not suggest that integrating over an entire event results in the clearer emergence of quadrupolar or dipolar symmetry.

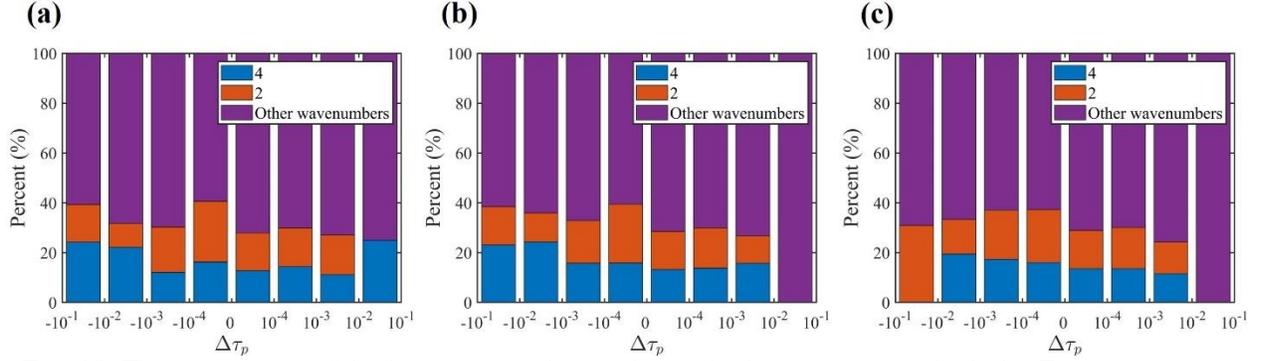

Fig. 12. The percentage of the long events, during the stick-slip regime, in which the Fourier spectrum, computed at discrete wavenumbers from 2 to 8 along a dashed circle with a radius of $8r_p$ surrounding the location where the shear strain is greatest, has the highest value at wavenumbers 4, 2, or others within each range of $\Delta\tau_p$ at (a) the first, (b) the second, and (c) the third timesteps.

3.3.2. Correlation analysis between macroscale response and kinematic measures

In the previous section, we observed that the clustering behavior in the GR and $GD^2_{min}$ regions occurs during most large slip events. This suggests that a single particle with the greatest $D^2_{min}$ value in a local region may play a key role in determining the overall particle rearrangement behavior. Consequently, $gD^2_{min}$ was identified as a parameter which may evolve in a correlated manner with $\Delta\tau$ across all timesteps of interest. On the other hand, the influence of an individual particle with the greatest $D^2_{min}$ value may extend beyond its immediate surroundings, potentially inducing shear transformations in neighboring zones depending on the local structural configuration. Therefore, $g_{t10\%}D^2_{min}$ was also introduced as an additional parameter for correlation analysis with $\Delta\tau$ across all timesteps.

Figure 13*a* illustrates the absolute values of the Spearman correlation coefficients between instantaneous values of $\Delta\tau$ and both $gD^2_{min}$ and $g_{t10\%}D^2_{min}$ across all timesteps for different bins of $\Delta\tau$ in the stick-slip regime. To ensure statistical significance, we exclude correlation coefficient data from all figures in this paper where the *P*-value exceeds 0.05. The analysis revealed that $\Delta\tau$ is generally weakly (correlation less than 0.4) to moderately (correlation between approximately 0.4 and 0.6) correlated with both $gD^2_{min}$ and $g_{t10\%}D^2_{min}$, respectively.

Since there is generally no strong correlation (greater than 0.6) between $\Delta\tau$ and instantaneous values of $gD^2_{min}$ or $g_{t10\%}D^2_{min}$ across all timesteps for different bins of $\Delta\tau$, we instead assessed the correlation between $\Delta\tau_p$, defined as the cumulative change in nondimensional shear stress from the start to the end of a stick or slip event, and two parameters: (1) the peak value of $gD^2_{min}$ during an event, as exemplified by step 4 in Figure 10b (Equation 2); (2) the peak value of $g_{t10\%}D^2_{min}$ during an event, observed at the same step in Figure 10b (Equation 3);

$$mgD^2_{min} = \max_{event}\left(gD^2_{min}\right) \tag{2}$$

$$mg_{t10\%}D^2_{min} = \max_{event}\left(g_{t10\%}D^2_{min}\right) \tag{3}$$

Figure 13b shows the absolute values of the Spearman correlation coefficients between $\Delta\tau_p$ and both $mgD^2_{min}$ and $mg_{t10\%}D^2_{min}$ for the stick and slip events for different bins of $\Delta\tau_p$ in the stick-slip regime. For the large slip events ($-10^{-1} < \Delta\tau_p < -10^{-3}$), the correlations between $\Delta\tau_p$ and both $mgD^2_{min}$ and $mg_{t10\%}D^2_{min}$ are moderate to strong, in contrast to the correlations between $\Delta\tau$ and both $gD^2_{min}$ and $g_{t10\%}D^2_{min}$, which are generally weak to moderate during slip events, as shown in Figure 13. These results indicate that the cumulative macroscale stress drop over an event, $\Delta\tau_p$, is closely associated with peak non-affine deformation during that event, while instantaneous macroscale stress drop, $\Delta\tau$, shows little correlation with concurrent non-affine activity.

The results also show that the correlation between $\Delta\tau_p$ and $mg_{t10\%}D^2_{min}$ is greater than that between $\Delta\tau_p$ and $mgD^2_{min}$ for the large slip events ($-10^{-1} < \Delta\tau_p < -10^{-3}$). This suggests that the variability in the $D^2_{min}$ values of the particles surrounding the particle with $gD^2_{min}$ affects $\Delta\tau_p$, supporting the notion that rearrangement events should be interpreted as non-local, regional phenomena, rather than point-like phenomena.

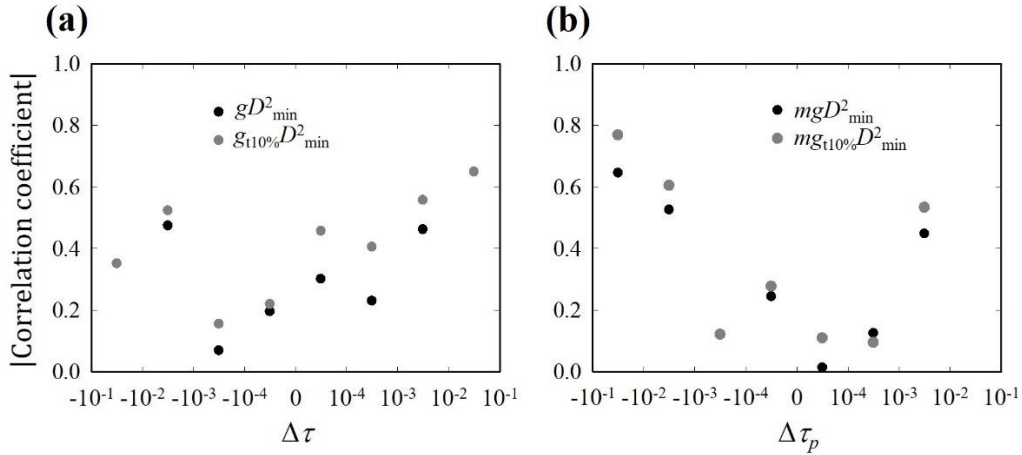

Fig. 13. Graphs of the absolute values of the Spearman correlation coefficients in the stick-slip regime: (a) between $\Delta\tau$ and both $gD^2_{min}$ and $g_{t10\%}D^2_{min}$, across all timesteps for different bins of $\Delta\tau$, (b) between $\Delta\tau_p$ and both $mgD^2_{min}$ and $mg_{t10\%}D^2_{min}$ for all stick and slip events for different bins of $\Delta\tau_p$. Correlation coefficients are not shown when the $p$ value exceeds 0.05.

To extend this analysis, we separated events into long and short categories, as defined in Section 3.3.1,

and examined how the correlations between $\Delta\tau_p$ and $mgD^2_{min}$, and between $\Delta\tau_p$ and $mg_{t10\%}D^2_{min}$ differ across event durations. Figure 14*a-b* shows the absolute values of Spearman correlation coefficients between $\Delta\tau_p$ and both $mgD^2_{min}$ and $mg_{t10\%}D^2_{min}$ for the long (*l*) events and the short (*s*) events, separately, in the stick-slip regime. In the long events, the correlations between $\Delta\tau_p$ and both $mgD^2_{min}$ and $mg_{t10\%}D^2_{min}$ during the slip events ($\Delta\tau_p < 0$) are all strong. Furthermore, consistent with Figure 13b, the correlation with $mg_{t10\%}D^2_{min}$ remains greater than that with $mgD^2_{min}$ for the large slip events ($-10^{-1} < \Delta\tau_p < -10^{-3}$), even when the analysis is restricted to the long events. In contrast, during the short events, those correlations during the slip events are mostly weak to moderate. This result suggests that the greatest non-affine motion during long slip events is strongly correlated with macroscale stress change, but this correlation is not observed during short events, which mostly correspond to transient stress fluctuations following slip events. Finally, the correlations during stick events ($\Delta\tau_p > 0$) in both the long and the short events are mostly weak.

To study the effect of the dynamics of particles surrounding a particle with $gD^2_{min}$ on macroscopic stress drops, removing the effect of the particle with $gD^2_{min}$ itself, we measured the Spearman partial correlation coefficient between $\Delta\tau_p$ and $mg_{t10\%}D^2_{min}$ for the long events. The Spearman partial correlation coefficient between $\Delta\tau_p$ and $mg_{t10\%}D^2_{min}$, removing the effect of $gD^2_{min}$, was calculated as [43]:

$$\rho_{\Delta\tau_p, mg_{t10\%}D^2_{min} | gD^2_{min}} = \frac{\rho_{\Delta\tau_p, mg_{t10\%}D^2_{min}} - \rho_{\Delta\tau_p, gD^2_{min}} \cdot \rho_{gD^2_{min}, mg_{t10\%}D^2_{min}}}{\sqrt{1-\rho^2_{\Delta\tau_p, gD^2_{min}}} \sqrt{1-\rho^2_{gD^2_{min}, mg_{t10\%}D^2_{min}}}} \quad (4)$$

where $\rho_{\Delta\tau_p, mg_{t10\%}D^2_{min}}$ is the Spearman correlation coefficient between $\Delta\tau_p$ and $mg_{t10\%}D^2_{min}$, $\rho_{\Delta\tau_p, gD^2_{min}}$ is the Spearman correlation coefficient between $\Delta\tau_p$ and $gD^2_{min}$, and $\rho_{gD^2_{min}, mg_{t10\%}D^2_{min}}$ is the Spearman correlation coefficient between $gD^2_{min}$ and $mg_{t10\%}D^2_{min}$. Even though $gD^2_{min}$ and $g_{t10\%}D^2_{min}$ generally follow the same trend during a slip event as shown in Figure 10b, time steps at which $mgD^2_{min}$ and $mg_{t10\%}D^2_{min}$ are measured can be slightly different within an event. Thus, when calculating the partial correlation coefficient in Eq. (4), we used the $gD^2_{min}$ value at the time step where $g_{t10\%}D^2_{min}$ is at its maximum during an event. Figure 14c shows these partial correlation coefficients. The results indicate that the partial correlations between $\Delta\tau_p$ and $mg_{t10\%}D^2_{min}$ for the long events are strong or moderate during the slip events ($\Delta\tau_p < 0$). This finding supports our previous argument that an STZ during slip events is regional rather than a point-like phenomenon.

From these partial correlation results, we found that nonaffine motion of neighboring particles around the one with $gD^2_{min}$, even when removing the direct contribution of particle with $gD^2_{min}$, is highly correlated with the cumulative nondimensional shear stress change. To further investigate the spatial extent of nonaffine motion and its correlation with stress change, we increased the $D^2_{min}$ cutoff radius from $5r_p$ to $10r_p$ and recalculated $mgD^2_{min}$ and $mg_{t10\%}D^2_{min}$, which are represented as $mgD^2_{min,10r_p}$ and $mg_{t10\%}D^2_{min,10r_p}$, respectively. Interestingly, as shown in Figure 14a, increasing the cutoff radius from $5r_p$ to $10r_p$ did not lead to a significant increase in the correlation between $\Delta\tau_p$ and both $mgD^2_{min}$ and $mg_{t10\%}D^2_{min}$ during the slip events ($\Delta\tau_p < 0$). Instead, the correlation between $\Delta\tau_p$ and $mg_{t10\%}D^2_{min}$ was significantly higher than that between $\Delta\tau_p$ and $mgD^2_{min}$. This indicates that measuring nonaffine motion within a larger radius around a particle with $gD^2_{min}$ does not effectively capture how particle rearrangements propagate from the $gD^2_{min}$ particle. Simply expanding the cutoff radius averages nonaffine motion over a broader region, causing the detailed propagation characteristics of rearrangements to be obscured. Conversely, considering only the top 10% nonaffine motions allows for effective identification of nonaffine motions of surrounding particles through which rearrangements propagate, thus better reflecting the actual propagation pattern and its correlation with cumulative shear stress changes.

Figure 14d shows the Spearman correlation coefficients between $\Delta\tau_p$ and $t_p$ for the long events and the short events, respectively. Consistent with the trends observed in Figures 11d and 11e, $\Delta\tau_p$ and $t_p$ are strongly correlated during the long and large stick events ($\Delta\tau_p > 10^{-3}$) but weakly correlated during the long slip events ($\Delta\tau_p < 0$). The short events of both types show weak to moderate correlations.

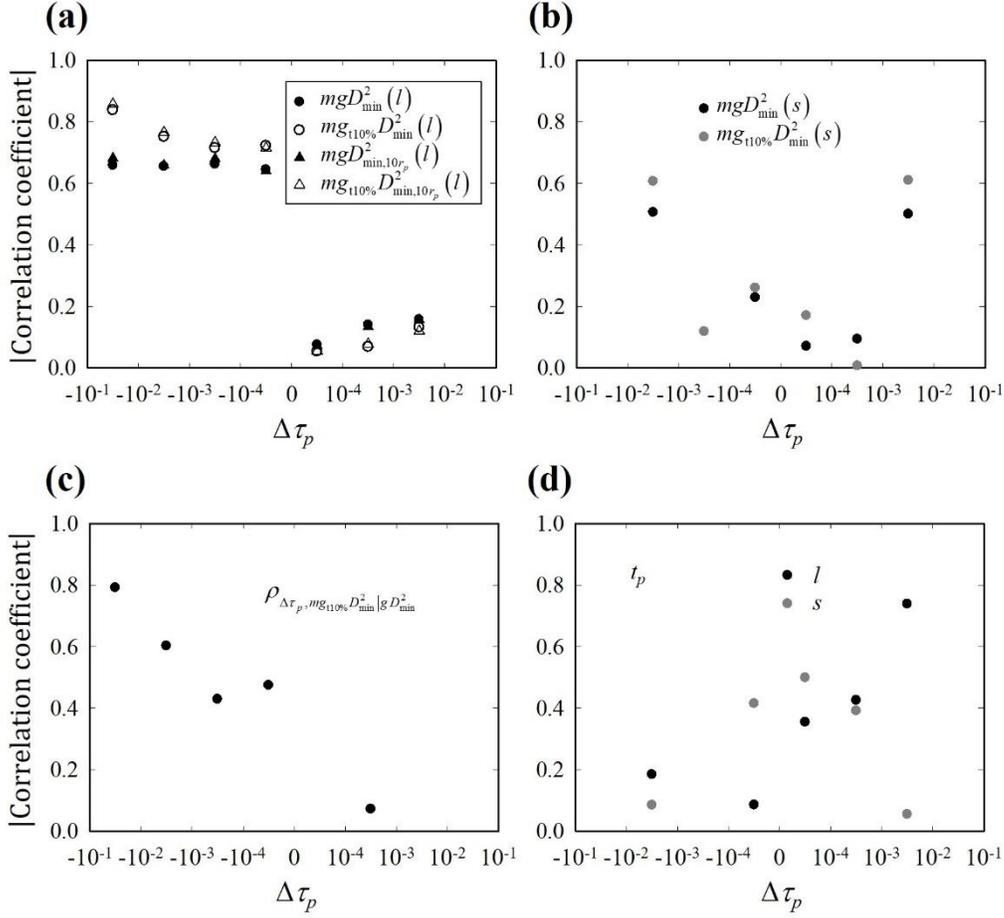

Fig. 14. Absolute values of the Spearman correlation coefficients for different bins of $\Delta\tau_p$ in the stick-slip regime: (a) between $\Delta\tau_p$ and $mgD^2_{\min}$, $mg_{t10\%}D^2_{\min}$, $mgD^2_{\min,10r_p}$, and $mg_{t10\%}D^2_{\min,10r_p}$ for the long (*l*) events, (b) between $\Delta\tau_p$ and both $mgD^2_{\min}$ and $mg_{t10\%}D^2_{\min}$ for the short (*s*) events, (c) Spearman partial correlation coefficients for $\Delta\tau_p$ with $mg_{t10\%}D^2_{\min}$ removing the effect of $gD^2_{\min}$ for the long events, and (d) Spearman correlation coefficient for $\Delta\tau_p - t_p$ for the long events and the short events, respectively.

In the Supplemental Materials (Section S3) [42], correlation results between $\Delta\tau_p$ and four additional parameters defined during an event are provided to offer additional context, while the main text focuses on the most representative measures discussed above. Correlation analyses involving the contact structure and clustering characteristics of the GR and $GD^2_{\min}$ regions are included as well; these results do not indicate any statistically significant relationship with $\Delta\tau_p$.

3.4. Other inter-particle friction coefficients

In Subsection 3.3.1, we conducted a statistical analysis to assess whether the representative characteristics of stick and slip events identified in earlier sections are consistently observed across multiple long events, confirming their typical features of such events. Furthermore, we found that the strain field around STZs does not clearly exhibit quadrupolar or dipolar symmetry during both long stick and long slip events. In Subsection 3.3.2, our correlation analyses examined the relationships between macroscale response and various kinematic measures, suggesting that an STZ behaves as a regional plastic event rather than a point-like event. Additionally, the correlation analyses investigated the effect of the microstructure of clustered GR and $GD^2_{min}$ regions on macroscale response.

We examined the observed statistical and correlation trends across different inter-particle sliding friction coefficients ($\mu_s$ = 0.1, 0.3, and 0.5). The detailed DEM results are provided in the Supplemental Material (Section S4) [42].

The results show that the statistical and correlation trends remain consistent across all values of the sliding friction coefficient. We initially expected that for low $\mu_s$, the positive correlation between $\Delta\tau_p$ and $t_s$ would not be significant due to reduced constraints between particles, which could lead to increased particle movement and rearrangement. This, in turn, could terminate a stick event and initiate a slip event. However, even for low $\mu_s$, the positive correlation between $\Delta\tau_p$ and $t_s$ remains significant (Figure S6). Given this, we further analyzed the correlation between $\Delta\tau_p$ and $t_s$ during the long and large stick events ($\Delta\tau_p > 10^{-3}$) and found that it remains strong across all values of $\mu_s$ (Figure S9h).

We hypothesized that $\mu_s$ could influence the features of shear strain field patterns around the STZ. At high $\mu_s$, strong particle constraints could restrict the long-range elastic propagation around the STZ, potentially suppressing the spatial development of quadrupolar or dipolar symmetry. In contrast, at low $\mu_s$, reduced constraints might allow the long-range elastic propagation around the STZ to occur more freely, facilitating the development of characteristic strain field patterns and leading to a more frequent observation of quadrupolar or dipolar symmetry in slip events. However, our results indicate that the occurrence of quadrupolar or dipolar symmetry does not increase at lower $\mu_s$. Instead, their occurrence remains nearly independent of $\mu_s$, as shown in Figure S7. This suggests that $\mu_s$ does not strongly influence the spatial development of long-range strain field patterns.

## 4. Discussion

In this study, we used a fixed small strain increment to capture the temporal evolution of stick and slip events with high resolution. This allowed us to identify spatial clustering of particles in the GR and $GD^2_{min}$ regions during large slip events and to analyze the associated dynamics in detail.

Our results show that the greatest non-affine motion, typically occurring within these clusters, strongly correlates with the magnitude of macroscale stress drops. This supports the view that these clusters represent shear transformation zones (STZs) whose activity governs macroscopic responses. Notably, only one or two clusters typically form during a large slip event, and the number of clusters does not significantly correlate with the stress drop magnitude. This indicates that it is the magnitude of local particle rearrangements rather than the number of clusters that governs the macroscale stress drop.

Although simple elastoplastic models (EPMs) predict that long-range elastic perturbations from a point-wise plastic event determine the surrounding strain field, our findings reveal a more complex picture. Even after excluding the particle with the greatest non-affine motion, a strong partial correlation remains between the macroscale stress drop and the mean non-affine motion of neighboring particles. This suggests that the stress redistribution process involves not only the most active site but also collective rearrangements in its vicinity.

Furthermore, while point-like plastic events are theoretically expected to induce quadrupolar or dipolar strain fields, our statistical analysis shows that such symmetry does not emerge as a statistically prominent feature. This implies that the strain field is not solely governed by elastic propagation from a point source but is shaped by regional plasticity and short-range rearrangements. Taken together, our observations indicate that plastic events in granular materials exhibit inherently regional characteristics, challenging simplified views based on point-like assumptions.

## 5. Conclusions

Our analysis of stick-slip events in granular materials reveals that plastic events are not point-like but regionally distributed, typically involving spatial clusters of particles in the GR and $GD^2_{min}$. The greatest non-affine motion within these clusters correlates strongly with macroscale stress drops, supporting their role as active zones of deformation.

We also find that the surrounding particle rearrangements contribute significantly to stress redistribution, even beyond the most active site. Additionally, the expected quadrupolar or dipolar symmetry in strain fields does not emerge as a statistically prominent feature, suggesting that strain fields are shaped by both short-range interactions and long-range elastic effects, with our results highlighting the role of collective dynamics and local rearrangements.

While we hypothesized that sliding friction might influence the emergence of long-range strain patterns, our results do not support a strong dependence on friction. The mechanisms governing the transition between long-range and short-range behavior remain unclear.

Future work should focus on the role of force chain evolution in stress redistribution. In particular,

investigating how the restructuring of force networks affects plastic activity during slip events, and how the stability of force chains relates to local rearrangements, may provide deeper insights into the mechanisms of strain localization in disordered materials.

**Acknowledgements**

This research was supported by the U.S. National Science Foundation CAREER Grant No. CBET-1942096. The authors also gratefully acknowledge partial support from the U.S. Geological Survey Earthquake Hazards Program through Award No. G23AP00268.

**Appendix A. Contact Models**

The normal contact force, $F_n$, consists of the normal elastic contact force, $F_n^e$, and the normal damping force, $F_n^d$, as

$$F_n = F_n^e + F_n^d = k_n \delta_n + \gamma_n \dot{\delta}_n \tag{A1}$$

where $k_n$ is the normal contact stiffness, $\delta_n$ is the normal overlapped displacement, $\gamma_n$ is the viscoelastic damping constant for the normal contact force, and $\dot{\delta}_n$ is the normal relative velocity between the two particles. The $k_n$ for cylindrical bodies can be calculated as [33]

$$k_n = \pi E^* L \left[ \log(4R_1/a) + \log(4R_2/a) - 1 \right]^{-1} \tag{A2}$$

$$\frac{1}{E^*} = \frac{1-\nu_1^2}{E_1} + \frac{1-\nu_2^2}{E_2} \tag{A3}$$

where $L$ is the cylinder length, $R_1$ and $R_2$ are the radius of particles 1 and 2, respectively, $a$ is the radius of contact area, $E_1$ and $E_2$ are Young's modulus of particles 1 and 2, respectively, and $\nu_1$ and $\nu_2$ are Poisson's ratio of particles 1 and 2, respectively. By using the linear damping model [35], the $\gamma_n$ can be calculated as

$$\gamma_n = 2\alpha \sqrt{k_n m^*} \tag{A4}$$

$$\frac{1}{m^*} = \frac{1}{m_1} + \frac{1}{m_2} \tag{A5}$$

where $\alpha$ is the damping ratio without dimension, and $m_1$ and $m_2$ are the mass of particles 1 and 2, respectively.

The tangential contact force, $F_t$, consists of the tangential elastic contact force, $F_t^e$, and the tangential damping force, $F_t^d$, as

$$F_t = F_t^e + F_t^d \tag{A6}$$

To consider the effects of various normal contact forces, $F_t^e$ is incrementally calculated as [34, 44]

$$F_t^e = F_t^0 + k_t \Delta \delta_t \tag{A7}$$

where $F_t^0$ is the tangential contact force at the previous time step, $k_t$ is the tangential contact stiffness, $\delta_t$ is the tangential overlapped displacement, and $\Delta \delta_t$ is the incremental tangential overlapped displacement. By using Lai *et al.*'s semi-analytical Hertzian frictional contact model in two dimensions [34], $k_t$ can be calculated as

$$k_t = \pi E^* L \Big[ \log(2R_1/a) + \log(2R_2/a) + 1 + w_1^{(a)} \big( \log(4R_1/a) + v_1/2(1-v_1) \big) \\ + w_2^{(a)} \big( \log(4R_2/a) + v_2/2(1-v_2) \big) \Big]^{-1} \tag{A8}$$

Here, $w_1^{(a)}$ and $w_2^{(a)}$ are the penalty factors, which can be calculated as

$$w_1^{(a)} = 0.22 + 2.89(a/R_1) + 0.18v_1 - 10.31(a/R_1)^2 + 1.75(a/R_1)v_1 + 0.36v_1^2 \\ w_2^{(a)} = 0.22 + 2.89(a/R_2) + 0.18v_2 - 10.31(a/R_2)^2 + 1.75(a/R_2)v_2 + 0.36v_2^2 \tag{A9}$$

The tangential elastic contact force, $F_t^e$, calculated in Eq. (A7) is subjected to Coulomb's law of friction as

$$F_t^e = \min\big(F_t^0 + k_t \Delta \delta_t, \mu_s F_n\big) \tag{A10}$$

where $\mu_s$ is the sliding friction coefficient.

The tangential damping force, $F_t^d$, is calculated as

$$F_t^d = \gamma_t \dot{\delta}_t \tag{A11}$$

where $\gamma_t$ is the viscoelastic damping constant for tangential force and $\dot{\delta}_t$ is the tangential relative velocity between the two particles. By using the linear damping model [35], the $\gamma_t$ can be calculated as

$$\gamma_t = 2\alpha \sqrt{k_t m^*} \tag{A12}$$

In the type C elasto-plastic spring-dashpot models [36], the total rolling resistance torque, $M_r$, consists of a spring torque, $M_r^k$, and a viscous damping torque, $M_r^d$, as

$$M_r = M_r^k + M_r^d \tag{A13}$$

This model can be implemented in a computer program in an incremental manner. The incremental spring torque, $\Delta M_r^k$, is calculated from the incremental relative rotation between two particles, $\Delta\theta_r$, and the rolling stiffness, $k_r$, as

$$\Delta M_r^k = -k_r \Delta\theta_r \tag{A14}$$

where $k_r$ is calculated as [45]

$$k_r = \frac{\pi E^* R_r^2 L}{4} \tag{A15}$$

$$\frac{1}{R_r} = \frac{1}{R_1} + \frac{1}{R_2} \tag{A16}$$

If the spring torque at time $t$ is $M_{r,t}^k$, the spring torque at time $t + \Delta t$ is given as

$$\begin{cases} M_{r,t+\Delta t}^k = M_{r,t}^k + \Delta M_r^k \\ \left| M_{r,t+\Delta t}^k \right| \leq M_r^m \end{cases} \tag{A17}$$

where $M_r^m$ is the limiting spring torque calculated as

$$M_r^m = \mu_r R_r F_n \tag{A18}$$

Here, $\mu_r$ is the rolling friction coefficient.

The viscous damping torque $M_r^d$ is assumed to be dependent on the relative rolling angular velocity, $\Delta\dot{\theta}_r$, between the two particles in contact and the rolling viscous damping constant, $C_r$, such that

$$M_{r,t+\Delta t}^d = \begin{cases} -C_r \dot{\theta}_r & \text{if } \left| M_{r,t+\Delta t}^d \right| < M_r^m \\ 0 & \text{if } \left| M_{r,t+\Delta t}^k \right| = M_r^m \end{cases} \tag{A19}$$

The $C_r$ is expressed as

$$C_r = \eta_r C_r^{\text{crit}} \tag{A20}$$

where $\eta_r$ is the rolling viscous damping ratio and $C_r^{\text{crit}}$ is the rolling critical viscous damping constant calculated as

$$C_r^{\text{crit}} = 2\sqrt{I_r k_r} \tag{A21}$$

in which $I_r$ is the equivalent moment of inertia for the relative rotational vibration mode about the contact point between the two contacting disks expressed as

$$\frac{1}{I_r} = \frac{1}{I_1 + m_1 R_1^2} + \frac{1}{I_2 + m_2 R_2^2} \tag{A22}$$

where $I_1$ and $I_2$ are the moments of inertia of particles 1 and 2, respectively, with respect to their own centroids.

In DEM simulation, to prevent an artificial jump in the elastic tangential overlapped displacement vector, $\delta_t$, $\delta_t$ is updated as [46]

$$\delta_t(t_i) = \begin{cases} \delta_t(t_{i-1}) + v_{c,t}(t_i)(t_i - t_{i-1}) & \text{when } \mathbf{F}_t(t_i) \le \mu\|\mathbf{F}_n(t_i)\| \\ \delta_t(t_{i-1}) - \dfrac{1}{k_t(t_i)}\left(\mathbf{t}\mu_s\|\mathbf{F}_n(t_i)\| - \mu\mathbf{F}_t(t_{i-1})\right) & \text{when } \mathbf{F}_t(t_i) > \mu\|\mathbf{F}_n(t_i)\| \end{cases} \tag{A23}$$

where $t_i$ and $t_{i-1}$ are the time steps at $i$ and $i$-1, respectively, $v_{c,t}$ is the tangential component of the relative velocity vector at the contact point of the particles, $\mathbf{t}$ is the unit vector of the tangential interaction, $F_n$ is the normal contact force vector, and $F_t$ is the tangential contact force vector.

### Appendix B. $D^2_{min}$ Computation

As described in the main text, $D^2_{min}$ is a measure of the magnitude of the local non-affine deformation. Specifically, $D^2_{min}$ quantifies the mean-square difference between the actual displacements of neighboring particles in the local averaging region relative to a central particle and the displacements that would occur if they were in a uniformly deformed region described by a local affine tensor, $\Gamma$. The local averaging regions are defined as regions of radius $r = Nr_p$, where $N$ is an integer, centered on the center of mass of a particle with mean radius $r_p$. Therefore, with a given strain increment at the scale of the system, $\Delta\gamma$, the $D^2_{min}$ for a specific particle $i$ can be calculated as [30]

$$D_{\min}^2(\gamma, \Delta\gamma) = \frac{1}{N_i} \sum_{j}^{N_i} \left\| \mathbf{r}_j(\gamma) - \mathbf{r}_i(\gamma) - \Gamma\left[\mathbf{r}_j(\gamma - \Delta\gamma) - \mathbf{r}_i(\gamma - \Delta\gamma)\right] \right\|^2 \tag{B1}$$

where $\|\cdot\|$ represents the L2 norm, $\gamma$ is the macroscopic shear strain of a given time, $N_i$ is the number of the neighbor particles of the particle $i$ within a local average region, and $j$ represents the neighboring particles of the particle $i$. At a given strain, $\gamma - \Delta\gamma$, particles $j$ that surround the particle $i$ are defined to form its neighborhood. $\mathbf{r}_i(\gamma)$ and $\mathbf{r}_j(\gamma)$ are the vectors from the origin of coordinate systems to the centers of the particles $i$ and $j$ at a given strain, $\gamma$, respectively. The locally best-fit affine tensor, $\Gamma$, can be obtained by minimizing the quantity $D^2_{\min}$ as [4]

$$\mathbf{X} = \sum_{j}^{N_i} \left[\mathbf{r}_j(\gamma) - \mathbf{r}_i(\gamma)\right] \otimes \left[\mathbf{r}_j(\gamma - \Delta\gamma) - \mathbf{r}_i(\gamma - \Delta\gamma)\right] \tag{B2}$$

$$\mathbf{Y} = \sum_{j}^{N_i} \left[\mathbf{r}_j(\gamma - \Delta\gamma) - \mathbf{r}_i(\gamma - \Delta\gamma)\right] \otimes \left[\mathbf{r}_j(\gamma - \Delta\gamma) - \mathbf{r}_i(\gamma - \Delta\gamma)\right] \tag{B3}$$

$$\Gamma = \mathbf{X} \cdot \mathbf{Y}^{-1} \tag{B4}$$

**Appendix C. Granular Lagrangian Strain and Shear Strain Field Computation**

In Section 3.2, we define the shear strain as the *xy* component of the granular Lagrangian strain. Here, we provide a detailed formulation of the granular Lagrangian strain used in our analysis.

The granular Lagrangian strain is formulated analogously to that of a continuous material, where the displacement fields are interpolated using finite element (FE) shape functions, as described for 3D in Zhang et al. [47]. Similarly, for 2D, the granular Lagrangian strain can be formulated as follows.

In 2D, since 3-noded triangular cells are constructed based on Delaunay tessellation, linear shape functions are employed to interpolate particle displacements. The spatial derivative of the current particle centroid displacement, *u*, with respect to the reference position, *X*, is then computed within an equivalent continuous domain, where the discrete particle system is treated as a continuum. This enables the displacement fields to be interpolated using linear shape functions, with triangular elements identified through Delaunay tessellation. We denote the three nodes of a triangle as A, B, and C. The position of node A in the reference configuration is given by $p^A = (p_x^A, p_y^A)$, and its nodal displacement is $u^A = (u_x^A, u_y^A)$. Using these notations, we define $p_x^{AB} = p_x^A - p_x^B$ and $u_x^{AB} = u_x^A - u_x^B$. Then, the displacement gradient is given as [48]:

$$\frac{\partial \boldsymbol{u}}{\partial \boldsymbol{X}} = \frac{1}{\det \mathbf{J}} \begin{Bmatrix} p_y^{BC} u_x^{AC} - p_y^{AC} u_x^{BC} & -p_x^{BC} u_x^{AC} + p_x^{AC} u_x^{BC} \\ p_y^{BC} u_y^{AC} - p_y^{AC} u_y^{BC} & -p_x^{BC} u_y^{AC} + p_x^{AC} u_y^{BC} \end{Bmatrix} \tag{C1}$$

where **J** is defined as [48]:

$$\mathbf{J} = \begin{Bmatrix} p_x^{AC} & p_y^{AC} \\ p_x^{BC} & p_y^{BC} \end{Bmatrix} \tag{C2}$$

Finally, the granular Lagrangian strain tensor at each triangular element is computed as:

$$\boldsymbol{E} = \frac{1}{2}\left[\left(\frac{\partial \boldsymbol{u}}{\partial \boldsymbol{X}}\right) + \left(\frac{\partial \boldsymbol{u}}{\partial \boldsymbol{X}}\right)^{\mathrm{T}} + \left(\frac{\partial \boldsymbol{u}}{\partial \boldsymbol{X}}\right)^{\mathrm{T}}\left(\frac{\partial \boldsymbol{u}}{\partial \boldsymbol{X}}\right)\right] \tag{C3}$$

The shear strain is then obtained as the *xy*-component of the granular Lagrangian strain tensor, *i.e.*, $E_{xy}$. To construct the shear strain field, we interpolate the shear strain values computed at the particle centroids onto a uniform grid. A linear interpolation method is used to estimate the strain values at the grid points. This approach ensures a smooth and continuous representation of the strain field over the domain.

# Supplemental Material: The Interplay Between Forces, Particle Rearrangements, and Macroscopic Stress Fluctuations in Sheared 2D Granular Media

Kwangmin Lee [a], Ryan C. Hurley [a, b] *

[a] *Department of Mechanical Engineering, Johns Hopkins University, Baltimore, Maryland 21218, USA*
[b] *Hopkins Extreme Materials Institute, Johns Hopkins University, Baltimore, Maryland 21218, USA*

**S1. Effect of Dashed Circle Radius on the Fourier Spectrum around the STZ**

In Section 3.2, a Fast Fourier Transform (FFT) was applied to analyze the angular wavenumber spectrum along dashed circles centered at the location of greatest shear strain in the sample, as shown in Figure 9. The dashed circles were initially set with a radius of $8r_p$.

In this section, we investigate how varying the radius of these dashed circles affects the Fourier spectrum. FFT was performed along dashed circles of different radii surrounding the greatest shear strain location for the representative long slip event analyzed in the main text.

Figure S1 shows the Fourier spectrum of angular wavenumbers along dashed circles around the STZ with different radii during the significant stress drop at steps 2–4, as indicated in Fig. 3a, with radii of (a) $6r_p$ and (b) $10r_p$. Similar to the results for a radius of $8r_p$ in Figure 9, these results also show that neither wavenumber 2 nor 4 is significantly dominant, and the spectrum is evenly distributed across different wavenumbers regardless of the dashed circle radii at the given time steps. This indicates that the shear strain field does not strictly conform to discrete quadrupolar or dipolar patterns. This supports our view in the main text that the strain field exhibits a broad and irregular spectrum rather than a consistent dipolar or quadrupolar symmetry.

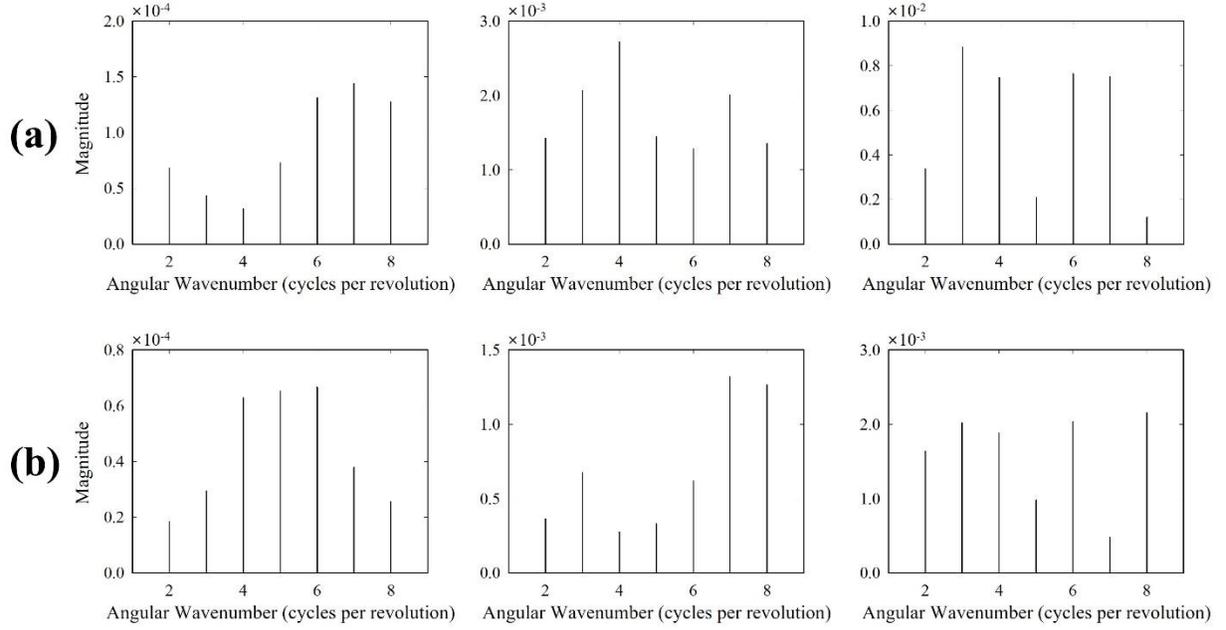

Fig. S1. Fourier spectrum of angular wavenumbers along the dashed circles with different radii with radii (a) $6r_p$ and (b) $10r_p$ during the significant stress drop at steps 2–4, as indicated in Fig. 3a. The columns from left to right correspond to steps 2, 3, and 4.

## S2. Effect of Time Interval Selection in $D^2_{min}$ Computation on Quadrupolar and Dipolar Symmetry

To examine whether the $D^2_{min}$ computation method affects the observed quadrupolar or dipolar symmetry, we recalculated $D^2_{min}$ using an alternative approach. Instead of using fixed short intervals, we integrated over entire stick or slip events. This method reduces local fluctuations, potentially enhancing the visibility of quadrupolar or dipolar symmetry.

To evaluate this effect, we performed FFT analysis along dashed circles of various radii centered at the location of greatest shear strain. The motivation for varying the radius is that recalculating $D^2_{min}$ over a longer time interval could expand the quadrupolar or dipolar symmetry region, as this approach uses a larger macroscale shear strain increment. Approximately 40% of the long events in the stick-slip regime were excluded from this statistical analysis because their greatest shear strain locations were near the sample boundary, which caused the dashed circles to extend beyond the sample boundary.

Figure S2 presents the percentage of the long events during the stick-slip regime where the Fourier spectrum, computed at discrete wavenumbers from 2 to 8 along dashed circles surrounding the location of greatest shear strain, has the highest value at wavenumbers 4, 2, or others. The dashed circles have radii of $8r_p$, $10r_p$, and $12r_p$, corresponding to (a), (b), and (c) in Figure S2. The results indicate that even with the recalculated $D^2_{min}$, the percentage of events showing dominant quadrupolar or dipolar symmetry remains low across all ranges of $\Delta\tau_p$. These findings indicate that the low occurrence of quadrupolar and dipolar symmetry in our study is likely a characteristic of the microscopic dynamics in the simulated system, rather than a consequence of the $D^2_{min}$ computation method.

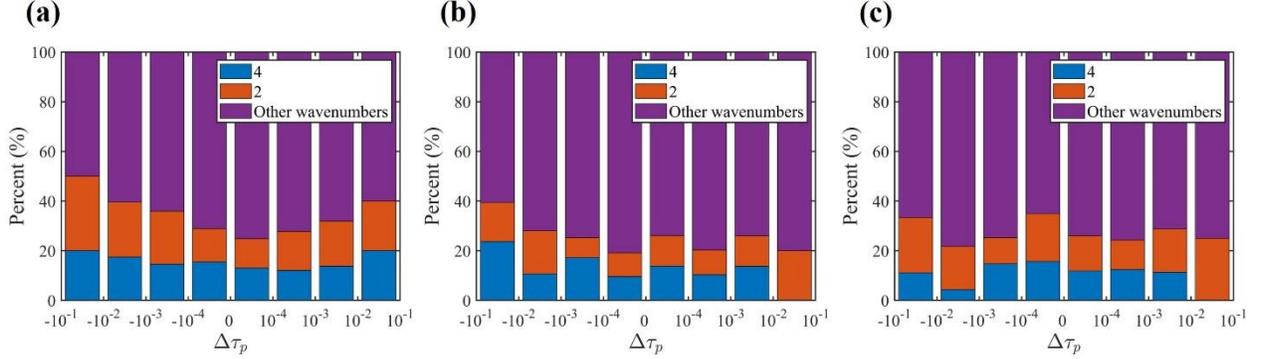

Fig. S2. The percentage of the long events during the stick-slip regime in which the Fourier spectrum, computed at discrete wavenumbers from 2 to 8 along dashed circles surrounding the location of greatest shear strain, has the highest value at wavenumbers 4, 2, or others. The dashed circles have radii of $8r_p$, $10r_p$, and $12r_p$, corresponding to (a), (b), and (c). The analysis is performed using the recalculated $D^2_{min}$ for each range of $\Delta\tau_p$ in the simulated stick-slip regime.

## S3. Additional Correlation Analyses for $\Delta\tau_p$

In section 3.3.2, we assessed the correlation between $\Delta\tau_p$ and two parameters. In this section, correlation results between $\Delta\tau_p$ and four additional parameters defined during an event are provided. The first two parameters are based on $gD^2_{min}$: (1) the value of $gD^2_{min}$ at $t_s$, the initial timestep when the clusters in the GR region are at their minimum during an event, as exemplified by step 2 in Figure 10c (Equation S1); (2) the sum of $gD^2_{min}$ values within an event, from $t_s$ to $t_e$, where $t_e$ is the concluding timestep of the stick or slip event, as exemplified by the timestep immediately following step 4 in the in Figure 10a (Equation S2). Mathematically, these parameters are given by:

$$gD^2_{min}(t_s) = gD^2_{min}\big|_{t=t_s} \tag{S1}$$

$$sgD^2_{min} = \sum_{t=t_s}^{t_e} gD^2_{min} \tag{S2}$$

Similarly, replacing $gD^2_{min}$ with $g_{t10\%}D^2_{min}$ in Equations (S1)-(S2), we define the third and fourth parameters as follows:

$$g_{t10\%}D^2_{min}(t_s) = g_{t10\%}D^2_{min}\big|_{t=t_s} \tag{S3}$$

$$sg_{t10\%}D^2_{min} = \sum_{t=t_s}^{t_e} g_{t10\%}D^2_{min} \tag{S4}$$

Figure S3a illustrates the absolute values of the Spearman correlation coefficients between $\Delta\tau_p$ and both $gD^2_{min}(t_s)$ and $g_{t10\%}D^2_{min}(t_s)$ for stick and slip events for different bins of $\Delta\tau_p$ in the stick-slip regime. The results show that $\Delta\tau_p$ is weakly to moderately correlated with both $gD^2_{min}(t_s)$ and $g_{t10\%}D^2_{min}(t_s)$. Notably, during slip events ($\Delta\tau_p < 0$), no strong correlation is observed. Therefore, the greatest non-affine motion near the shear transformation zone, initially generated during the significant stress drop, is not closely related to the magnitude of the stress drop during the slip events.

Figure S3b shows the absolute values of the Spearman correlation coefficients between $\Delta\tau_p$ and both $sgD^2_{min}$ and $sg_{t10\%}D^2_{min}$ for stick and slip events for different bins of $\Delta\tau_p$ in the stick-slip regime. Compared to the correlation trends shown in Figure 13b between $\Delta\tau_p$ and $mgD^2_{min}$ (or $mg_{t10\%}D^2_{min}$), the correlations in Figure S3b between $\Delta\tau_p$ and $sgD^2_{min}$ (or $sg_{t10\%}D^2_m$) are similar. Thus, considering $gD^2_{min}$ or $g_{t10\%}D^2_{min}$ in additional timesteps during a given event, beyond the timestep at which $mgD^2_{min}$ or $mg_{t10\%}D^2_{min}$ occurs, does not significantly change its correlation with $\Delta\tau_p$.

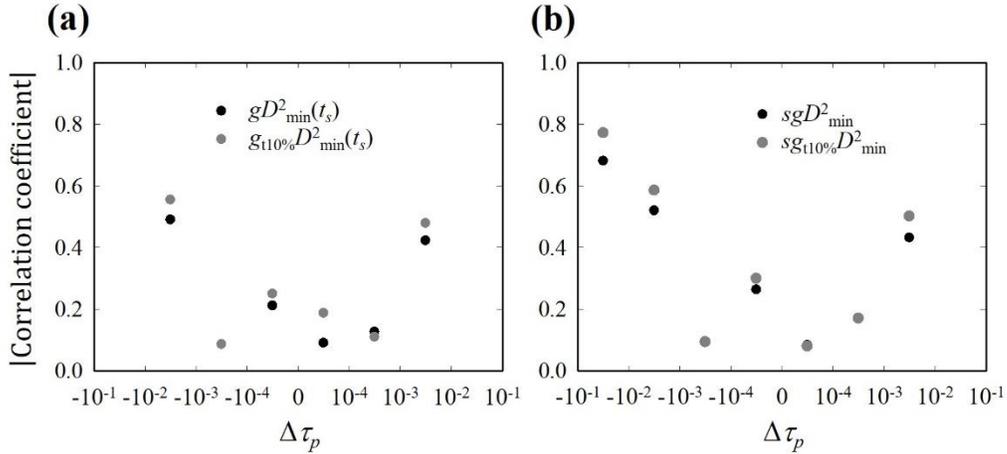

Fig. S3. Graphs of the absolute values of the Spearman correlation coefficients for all stick and slip events for different bins of $\Delta\tau_p$: (a) between $\Delta\tau_p$ and both $gD^2_{min}(t_s)$ and $g_{t10\%}D^2_{min}(t_s)$; (b) between $\Delta\tau_p$ and both $sgD^2_{min}$ and $sg_{t10\%}D^2_{min}$. Correlation coefficients are not shown when the $p$ value exceeds 0.05.

To study the correlation between $\Delta\tau_p$ and the microstructure of clustered GR region during slip events, we calculated the absolute values of Spearman correlation coefficients between $\Delta\tau_p$ and both the mean and the median of the absolute contact angles in the GR region for different bins of $mg_{t10\%}D^2_{min}$ during the long events in the stick-slip regime (Figure S4a). These correlations were calculated at the timestep $t_s$. The results indicate weak correlations or correlations lacking statistical significance ($P$-value > 0.05), suggesting that the contact structure of the single clustered GR region does not influence $\Delta\tau_p$. This is because $\Delta\tau_p$ is highly correlated with $mg_{t10\%}D^2_{min}$, but $mg_{t10\%}D^2_{min}$ occurs at a later timestep than $t_s$. For example, Figure 10d shows that the mean and the median of the contact angles at step 2 (at $t_s$) differ from those at step 4, where $mg_{t10\%}D^2_{min}$ occurs.

In section 3.3.1, we showed that the minimum number of clusters in the $GD^2_{min}$ and GR regions is typically one or two during large slip events. We investigated whether the minimum number of clusters in the $GD^2_{min}$ and GR regions correlates with $\Delta\tau_p$. Figure S4b shows the absolute values of Spearman correlation coefficients between $\Delta\tau_p$ and the minimum number of clusters in the $GD^2_{min}$ and GR regions, respectively, for different bins of $mg_{t10\%}D^2_{min}$ during the long events in the stick-slip regime. The results indicate weak correlations, or correlations lacking statistical significance ($P$-value > 0.05), suggesting that the minimum number of clusters in the $GD^2_{min}$ and GR regions has little effect on $\Delta\tau_p$.

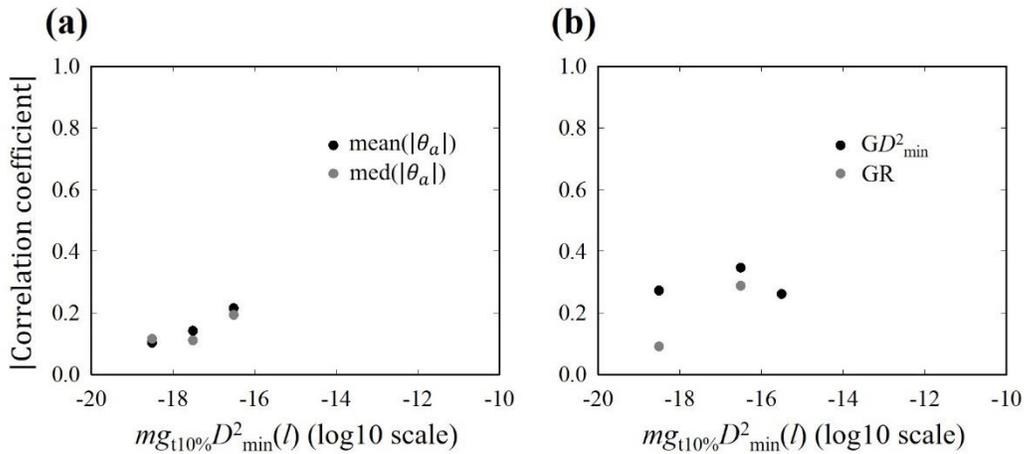

Fig. S4. (a) Absolute values of the Spearman correlation coefficients for different bins of $mg_{t10\%}D^2_{min}$ during the long events in the stick-slip regime: (a) between $\Delta\tau_p$ and both the mean and the median of the absolute contact angles in GR region at timestep $t_s$, and (b) between $\Delta\tau_p$ and the minimum number of clusters in $GD^2_{min}$ and GR region, respectively.

## S4. Statistical and Correlation Results of Stick-Slip Events for Various Sliding Friction Coefficients

To examine whether the statistical and correlation trends observed in the $\mu_s = 0.7$ simulation (Section 3.3) hold across different inter-particle sliding friction coefficients, we conducted additional DEM simulations with $\mu_s = 0.1$, 0.3, and 0.5. The boundary conditions, material properties, and contact properties remained the same as those used in the DEM simulation described in Section 2. Each simulation included approximately 130,000 timesteps, yielding approximately 26,000 events for $\mu_s = 0.1$, 23,000 for $\mu_s = 0.3$, and 25,000 for $\mu_s = 0.5$ within the stick-slip regime.

We investigated whether the clustering behavior in the GR and $GD^2_{min}$ regions is statistically significant across long slip events in the stick-slip regime for various sliding friction coefficients. In addition, we examined whether there is a positive correlation between $\Delta\tau_p$ and $t_p$ during long stick events across different sliding friction coefficients.

Figure S5 presents the statistics of the long events in the stick-slip regime for different inter-particle sliding friction coefficients. The first column shows the number of stick and slip events within each range of $\Delta\tau_p$. The second and third columns show the percentage of the minimum number of clusters during an event for different bins of $\Delta\tau_p$ in the GR and $GD^2_{min}$ regions, respectively. The first, second, and third rows correspond to $\mu_s = 0.1$, 0.3, and 0.5, respectively.

Figure S6 shows two-dimensional histograms of $\Delta\tau_p$ and $t_p$ for the long stick events and the long slip events, respectively, during the stick-slip regime. The results are shown for different inter-particle sliding friction coefficients: the first, second, and third rows correspond to $\mu_s = 0.1$, 0.3, and 0.5, respectively.

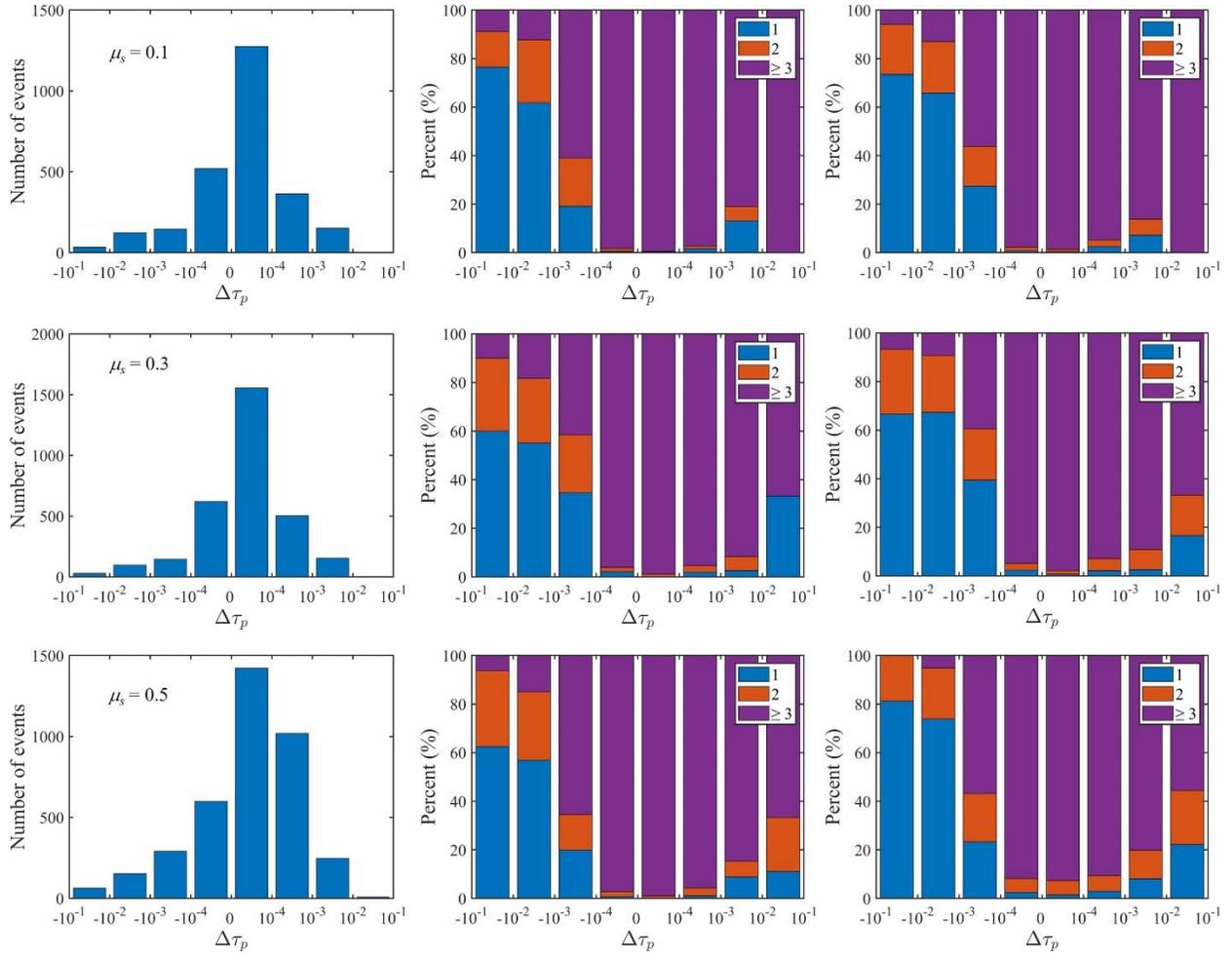

Fig. S5. Statistics of the long events in the stick-slip regime for different inter-particle sliding friction coefficients. The first column shows the number of stick and slip events within each range of $\Delta \tau_p$. The second and third columns show the percentage of the minimum number of clusters during an event for different bins of $\Delta \tau_p$ in the GR and $GD^2_{min}$ regions, respectively. The first, second, and third rows correspond to $\mu_s$ = 0.1, 0.3, and 0.5, respectively.

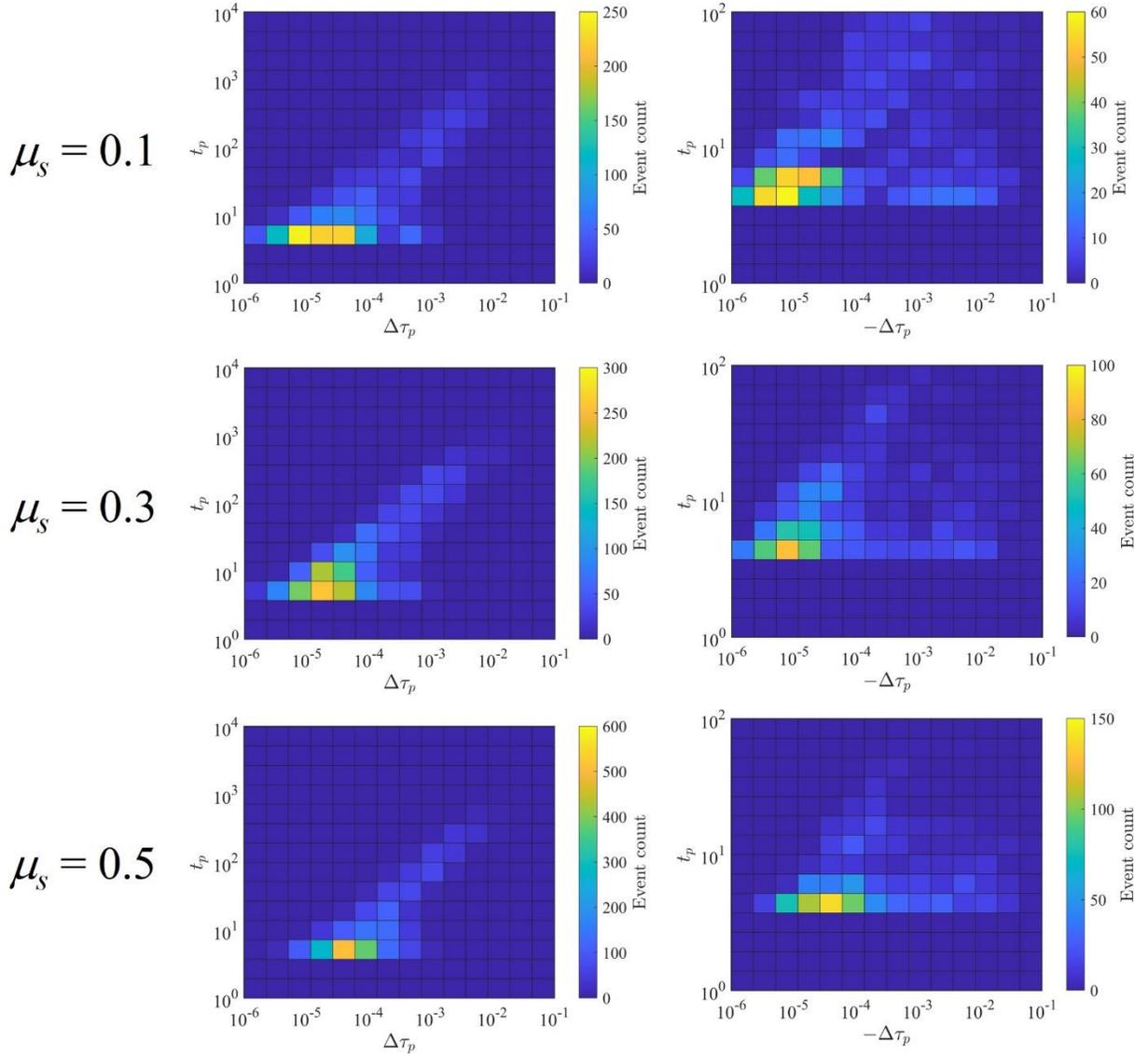

Fig. S6. Two-dimensional histograms showing the distribution of $\Delta\tau_p$ and $t_p$ for the long events during the stick-slip regime. The left and right columns correspond to the stick events and the slip events, respectively, and each row represents a different inter-particle sliding friction coefficient ($\mu_s = 0.1$, 0.3, and 0.5 from top to bottom). The color represents the number of events in each bin.

To examine the statistical behavior of non-affine motion for various sliding friction coefficients, we performed FFT along a dashed circle with a radius of $8r_p$ surrounding the location where the shear strain is greatest, as in Figure 9, for the long events during the stick-slip regime.

Figure S7 illustrates the percentage of the long events in the stick-slip regime in which the Fourier spectrum (computed at discrete wavenumbers from 2 to 8) has the highest value at wavenumbers 4, 2, or others within each range of $\Delta\tau_p$ for various friction coefficients. The results are shown for the first, the second, and the third timesteps, corresponding to the first, second, and third columns, respectively. The first, second, and third rows correspond to $\mu_s = 0.1$, 0.3, and 0.5, respectively.

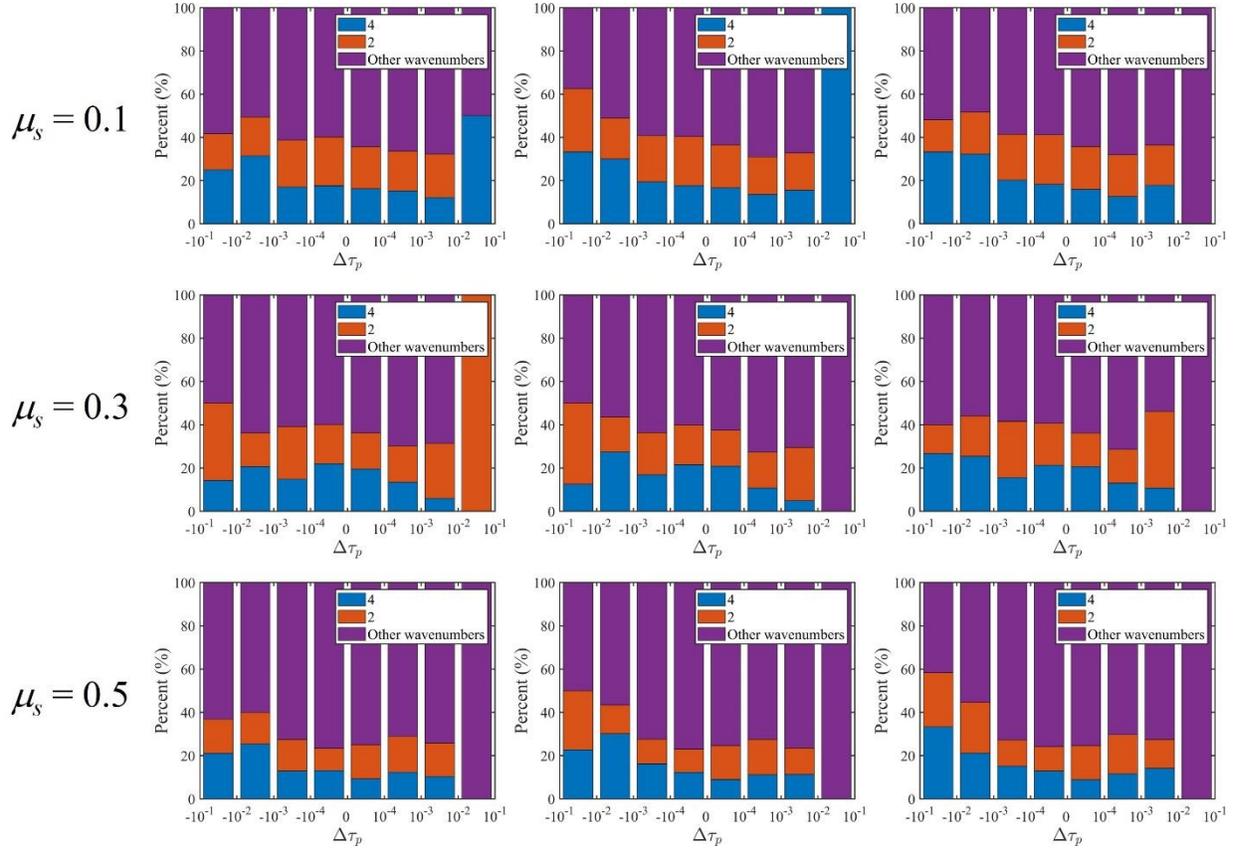

Fig. S7. The percentage of the long events in the stick-slip regime in which the Fourier spectrum has the highest value at wavenumbers 4, 2, or others within each range of $\Delta\tau_p$, across various friction coefficients. The spectrum is computed at discrete wavenumbers from 2 to 8 along a dashed circle with a radius of $8r_p$ surrounding the location where the shear strain is greatest. The first, the second, and the third timesteps correspond to the first, second, and third columns, respectively. The first, second, and third rows correspond to $\mu_s = 0.1$, 0.3, and 0.5, respectively.

We examined the correlations between various parameters and $\Delta\tau$ across all timesteps or $\Delta\tau_p$ for all events in the stick-slip regime.

Figure S8 illustrates the absolute values of the Spearman correlation coefficients between kinematic measures and $\Delta\tau$ or $\Delta\tau_p$ in the stick-slip regime for various sliding friction coefficients, with each value of $\mu_s$ represented by different symbols. Panels (a) and (b) show the correlation coefficients between $\Delta\tau$ and two measures, $gD^2_{min}$ and $g_{t10\%}D^2_{min}$, across all instantaneous timesteps. Panels (c) through (h) present the correlation coefficients between $\Delta\tau_p$ and six different kinematic measures: $gD^2_{min}(t_s)$, $g_{t10\%}D^2_{min}(t_s)$, $mgD^2_{min}$,

$mg_{t10\%}D^2_{min}$, $sgD^2_{min}$, and $sg_{t10\%}D^2_{min}$, for all events.

Figure S9 illustrates the absolute values of the Spearman correlation coefficients between kinematic measures and $\Delta\tau_p$ in the stick-slip regime for various sliding friction coefficients, with each value of $\mu_s$ represented by different symbols. Panels (a) to (f) present the correlation coefficients between $\Delta\tau_p$ and six different kinematic measures: $mgD^2_{min}(l)$, $mg_{t10\%}D^2_{min}(l)$, $mgD^2_{min,10r_p}(l)$, $mg_{t10\%}D^2_{min,10r_p}(l)$, $mgD^2_{min}(s)$, and $mg_{t10\%}D^2_{min}(s)$. Panel (g) presents the partial correlation coefficient between $\Delta\tau_p$ and $mg_{t10\%}D^2_{min}$ removing the effect of $gD^2_{min}$ for the long events. Panel (h) and (k) show the correlation coefficient between $\Delta\tau_p$ and $t_p$ for the long events and the short events, respectively.

Figure S10 illustrates the absolute values of Spearman correlation coefficients between $\Delta\tau_p$ and microstructure of clustered GR or $GD^2_{min}$ regions for different bins of $mg_{t10\%}D^2_{min}$ during the long events in the stick-slip regime. Panels (a) and (b) present the correlation between $\Delta\tau_p$ and both the mean and the median of the absolute contact angles in the GR regions, respectively. Panels (c) and (d) show the correlation between $\Delta\tau_p$ and the minimum number of clusters in the $GD^2_{min}$ region and the GR region, respectively.

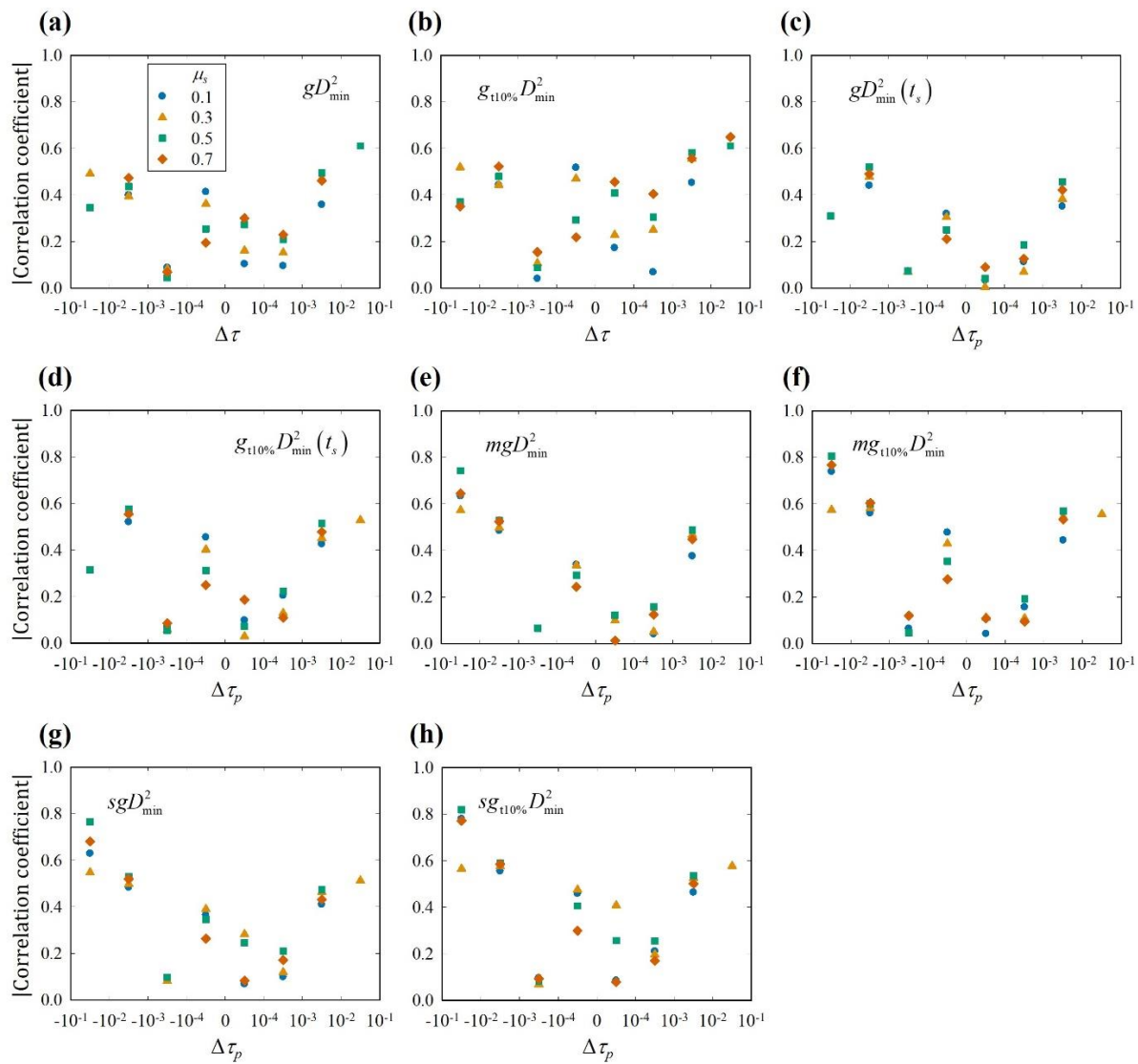

Fig. S8. The absolute values of the Spearman correlation coefficients between kinematic measures and $\Delta\tau$ or $\Delta\tau_p$ in the stick-slip regime for various sliding friction coefficients, with each value of $\mu_s$ represented by different symbols. Panels (a) and (b) show the correlation coefficients between $\Delta\tau$ and two measures, $gD^2_{min}$ and $g_{t10\%}D^2_{min}$, across all instantaneous timesteps. Panels (c) through (h) present the correlation coefficients between $\Delta\tau_p$ and six different kinematic measures: $gD^2_{min}(t_s)$, $g_{t10\%}D^2_{min}(t_s)$, $mgD^2_{min}$, $mg_{t10\%}D^2_{min}$, $sgD^2_{min}$, and $sg_{t10\%}D^2_{min}$, for all events.

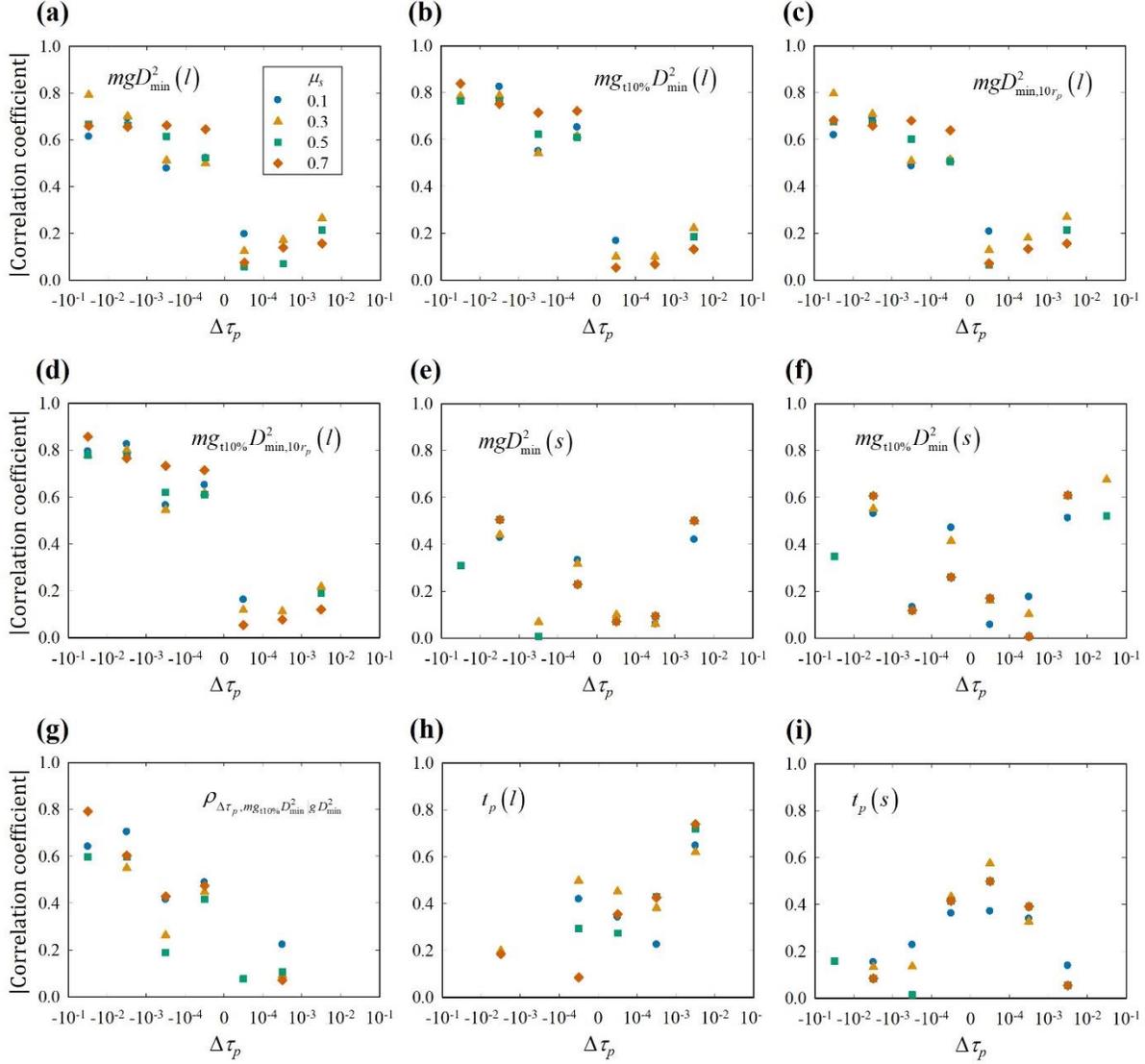

Fig. S9. The absolute values of the Spearman correlation coefficients between kinematic measures and $\Delta\tau_p$ in the stick-slip regime for various sliding friction coefficients, with each value of $\mu_s$ represented by different symbols. Panels (a) to (f) present the correlation coefficients between $\Delta\tau_p$ and six different kinematic measures: $mgD^2_{min}(l)$, $mg_{t10\%}D^2_{min}(l)$, $mgD^2_{min,10r_p}(l)$, $mg_{t10\%}D^2_{min,10r_p}(l)$, $mgD^2_{min}(s)$, and $mg_{t10\%}D^2_{min}(s)$. Panel (g) presents the partial correlation coefficient between $\Delta\tau_p$ and $mg_{t10\%}D^2_{min}$ removing the effect of $gD^2_{min}$ for the long events. Panel (h) and (k) shows the correlation coefficient between $\Delta\tau_p$ and $t_p$ for the long events and the short events, respectively.

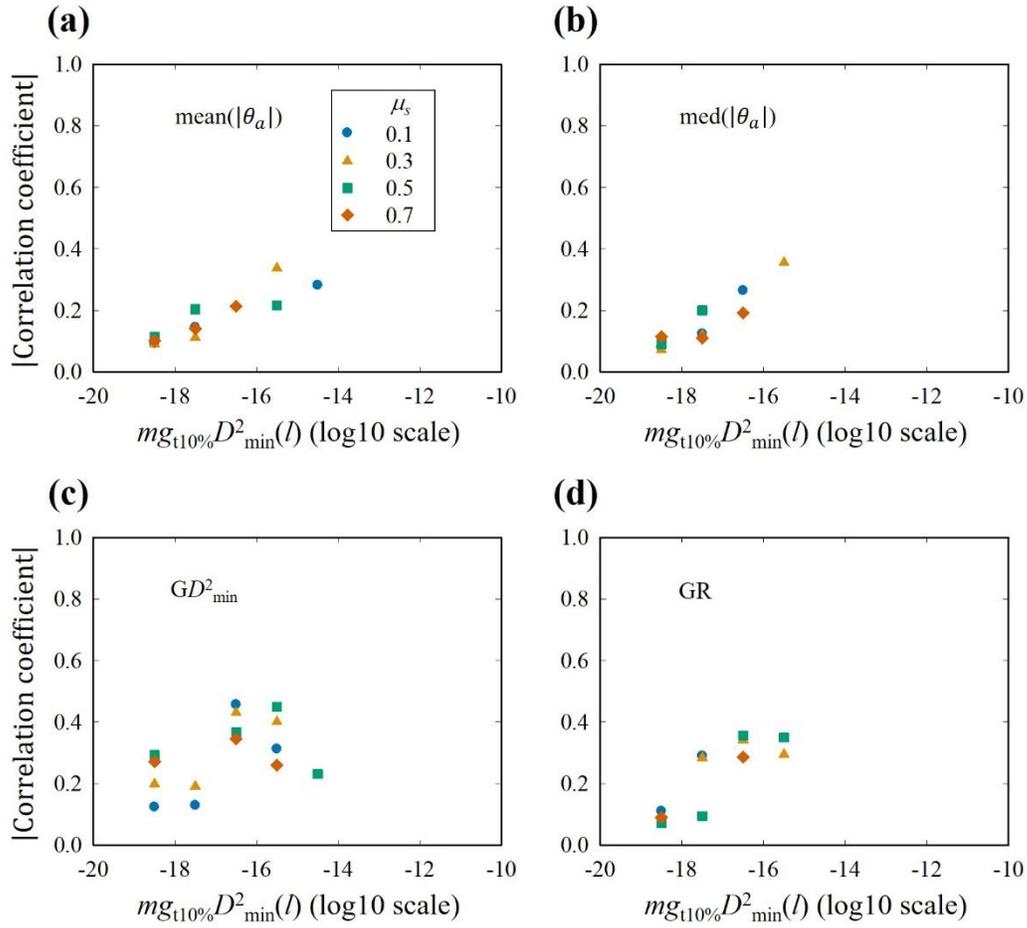

Fig. S10. The absolute values of Spearman correlation coefficients between $\Delta\tau_p$ and microstructure of clustered GR or $GD^2_{min}$ regions for different bins of $mg_{t10\%}D^2_{min}$ during the long events in the stick-slip regime. Panels (a) and (b) present the correlation between $\Delta\tau_p$ and both the mean and the median of the absolute contact angles in the GR regions, respectively. Panels (c) and (d) show the correlation between $\Delta\tau_p$ and the minimum number of clusters in the $GD^2_{min}$ region and the GR region, respectively.

13